\documentclass[prologue,dvipsnames,table,xcdraw, manuscript, sigconf, square]{acmart}

\AtBeginDocument{%
  \providecommand\BibTeX{{%
    \normalfont B\kern-0.5em{\scshape i\kern-0.25em b}\kern-0.8em\TeX}}}

\acmYear{2025}
\setcopyright{cc}
\setcctype{by}
\acmConference[SCF '25]{ACM Symposium on Computational Fabrication}{November 20--21, 2025}{Cambridge, MA, USA}
\acmBooktitle{ACM Symposium on Computational Fabrication (SCF '25), November 20--21, 2025, Cambridge, MA, USA}
\acmDOI{10.1145/3745778.3766650}
\acmISBN{979-8-4007-2034-5/2025/11}

\usepackage[ruled]{algorithm2e} % For algorithms

\SetAlFnt{\small}
\SetAlCapFnt{\small}
\SetAlCapNameFnt{\small}
\SetAlCapHSkip{0pt}

\acmJournal{TOG}

\usepackage{longtable}
\usepackage{listings}
\usepackage{color, soul}
\usepackage{subfig}
\usepackage{float}
\floatstyle{plaintop}
\restylefloat{table}
\usepackage{colortbl}
\usepackage[utf8]{inputenc}
\usepackage[T1]{fontenc}
\usepackage{enumitem}
\usepackage{siunitx}
\usepackage{makecell}
\usepackage{pifont}
\usepackage{verbatim}
\usepackage{multirow}
\usepackage{multicol}
\begin{document}

%% ------------------------------------------------------
%% Title - parameter in brackets is the short title
%% ------------------------------------------------------
\title{Computational Design and Single-Wire Sensing of 3D Printed Objects with Integrated Capacitive Touchpoints}

%% ------------------------------------------------------
%% Authorship
%% ------------------------------------------------------
\author{S. Sandra Bae}
\authornote{These authors contributed equally to this work and are listed alphabetically.}
\email{sandrabae@cs.arizona.edu}
\affiliation{
  \department{ATLAS Institute}    
  \institution{University of Colorado}
  \city{Boulder, CO}
  \country{USA}
}
\affiliation{
  \department{\& Dept. of Computer Science}    
  \institution{University of Arizona}
  \city{Tucson, AZ}
  \country{USA}
}

\author{Takanori Fujiwara}
\authornotemark[1]
\email{tfujiwara@cs.arizona.edu}
\affiliation{
  \department{Dept. of Science and Technology}    
  \institution{Link{\"o}ping University}
  \city{Link{\"o}ping}
  \country{Sweden}
}
\affiliation{
  \department{\& Dept. of Computer Science}    
  \institution{University of Arizona}
  \city{Tucson, AZ}
  \country{USA}
}

\author{Danielle Albers Szafir}
\email{danielle.szafir@cs.unc.edu}
\affiliation{
  \department{Dept. of Computer Science} 
  \institution{University of North Carolina}
  \city{Chapel Hill, NC}
  \country{USA}
}

\author{Ellen Yi-Luen Do}
\email{ellen.do@colorado.edu}
\affiliation{
  \department{ATLAS Institute}    
  \institution{University of Colorado}
  \city{Boulder, CO}
  \country{USA}
}

\author{Michael L. Rivera}
\email{mrivera@colorado.edu}
\affiliation{
  \department{ATLAS Institute \& Dept. of
Computer Science}    
  \institution{University of Colorado}
  \city{Boulder, CO}
  \country{USA}
}

\renewcommand{\shortauthors}{Bae and Fujiwara et al.}

\renewcommand{\sectionautorefname}{Sec.}
\renewcommand{\subsectionautorefname}{Sec.}
\renewcommand{\subsubsectionautorefname}{Sec.}
\renewcommand{\figureautorefname}{Fig.}
\newcommand{\subfigureautorefname}{Fig.}
\renewcommand{\tableautorefname}{Ta\-ble}
\renewcommand{\equationautorefname}{Eq.}

\begin{teaserfigure}
    \centering
    \includegraphics[width=\textwidth]{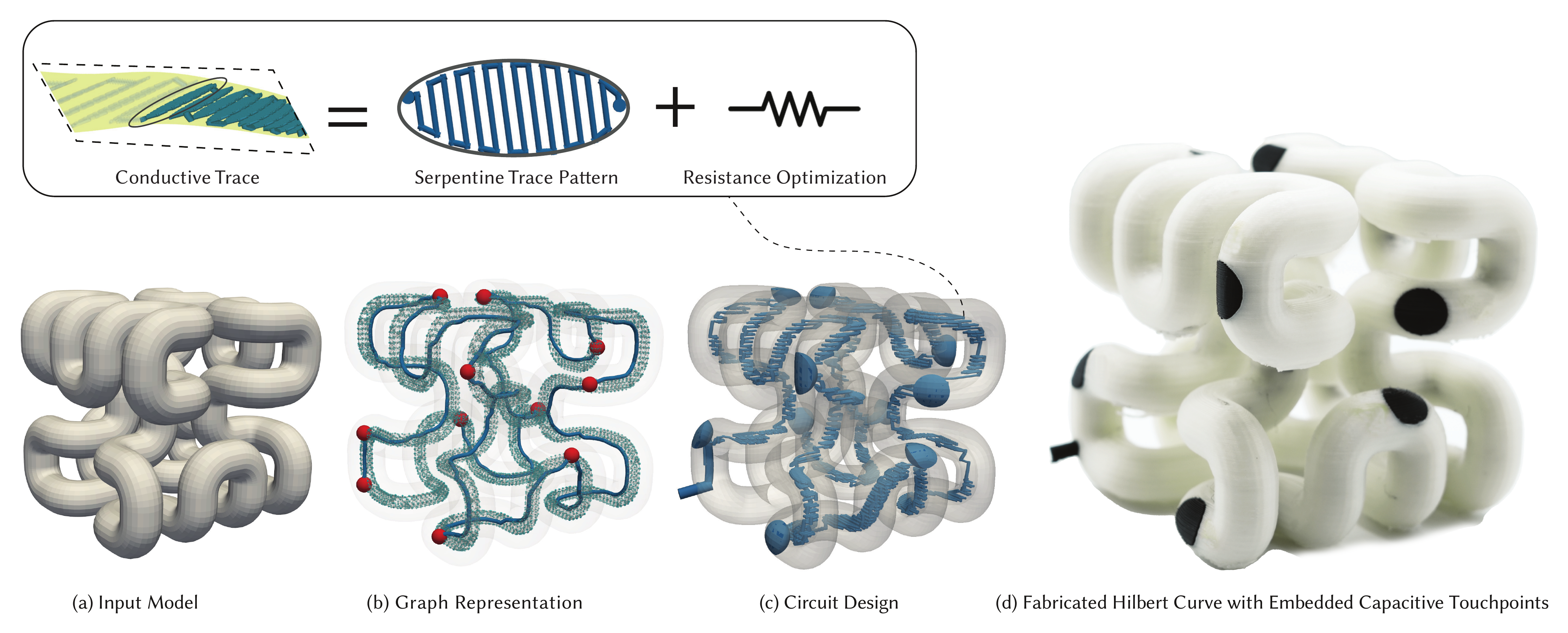}
    \caption{\textbf{High-level overview of our computational design pipeline:} (a) Our pipeline begins with an input model and the user selects different areas on the model's surface to turn into touchpoints; (b) computes a graph-based path to serially connect the touchpoints; (c) generates an internal circuit design to embed capacitive sensors inside the object; and (d) fabricates the object and internal circuit using multi-material 3D printing to be used as a sensing interface with only 1 or 2 wire connections. }
    \Description{
    Design process of illustrating how to integrate capacitive touch sensors to the Stanford Bunny from input model to graph representation, circuit design, and final fabricated product.
    }
    \label{fig:teaser}
\end{teaserfigure}

\def\singlewire{single-wire}
\def\doublewire{double-wire}
\def\points{points}
\def\pipes{conduits}
\def\pipe{conduit}
\newcommand{\nTouchpoints}{N}
\newcommand{\VIn}{v_\mathrm{in}}
\newcommand{\VThres}{v_\mathrm{thres}}
\newcommand{\TCalib}[1]{{t_\mathrm{cal}}_{#1}}
\newcommand{\TRecog}[1]{{t_\mathrm{rec}}_{#1}}
\newcommand{\RTill}[1]{{r_\mathrm{till}}_{#1}}
\newcommand{\RAfter}[1]{{r_\mathrm{after}}_{#1}}
\newcommand{\LogTerm}[1]{L_{#1}}

%% ------------------------------------------------------
%% Sections
%% ------------------------------------------------------
\begin{abstract}
Producing interactive 3D printed objects currently requires laborious 3D design and post-instrumentation with off-the-shelf electronics. Multi-material 3D printing using conductive PLA presents opportunities to mitigate these challenges.  We present a computational design pipeline that embeds multiple capacitive touchpoints into any 3D model that has a closed mesh without self-intersection. With our pipeline, users define touchpoints on the 3D object's surface to indicate interactive regions. Our pipeline then automatically generates a conductive path to connect the touch regions. This path is optimized to output unique resistor-capacitor delays when each region is touched, resulting in all regions being able to be sensed through a \doublewire\ or \singlewire\ connection. 
% Objects are then fabricated using multi-material 3D printing with conductive PLA.  
% We further extend this approach to a \singlewire\ connection. 
We illustrate our approach's utility with five computational and sensing performance evaluations (achieving 93.35\% mean accuracy for single-wire) and six application examples. Our sensing technique supports existing uses (e.g., prototyping) and highlights the growing promise to produce interactive devices entirely with 3D printing.
\end{abstract}


%   \includegraphics[width=\textwidth]{figures/Asset 1.pdf}
%   \caption{DIY Paper Charts using everyday materials and AR markers}


\maketitle
\section{Introduction}
\label{sec1:introduction}
Despite recent advances, the overall design and manufacturing process to fabricate interactive 3D printed objects is time-consuming and fragmented. Embedding off-the-shelf electronic components (e.g., sensors~\cite{zhu2020curveboards, wang2020morphingcircuit} or LEDs~\cite{he2022modelec, savage2014pipes}) into 3D prints is a popular approach, adhering to the traditional design process that separates form (i.e., designing a 3D model) and interactivity into two individual processes.  However, this approach introduces two challenges. 
First, it requires users to design \textit{around} the electronic components and their predefined shapes and dimensions. This constraint makes it difficult to integrate electronics into complex geometries, such as curved or organic shapes or thin-walled structures with limited bounding volume (e.g., sword, robotic tactile sensors~\cite{kohlbrenner2025procedural}).
Second, it requires users to have extensive knowledge spanning electronics, computer-aided design, and fabrication. Each step is compartmentalized to a dedicated software, such that a change requires modifying subsequent steps through extensive trial and error. Our work is motivated by the following question: \textit{how can we effectively streamline the process of manufacturing interactive 3D printed objects?} 

Multi-material printing can help address these challenges while presenting new design opportunities. 
Namely, it can bridge the two aforementioned processes (i.e., form, interactivity) into one streamlined process.
As one example, we can use conductive filaments to 3D print electronics, such as wires and resistors, directly into the target object and minimize post-instrumentation.
Minimizing instrumentation can enhance durability~\cite{zhu2020curveboards}, aesthetics~\cite{olberding2013cuttable, zhu2020curveboards}, and space efficiency~\cite{dahiya2009tactile} while simplifying (dis)assembly~\cite{he2022modelec, wen2025m2d2} and reducing costs~\cite{Rupavatharam2023ambi}.
This bridging can lead toward the broad vision of 3D printing objects that are fully interactive and ready to be used straight off the printer.

\begin{figure*}[t!]
    \centering
    \includegraphics[width=\linewidth]{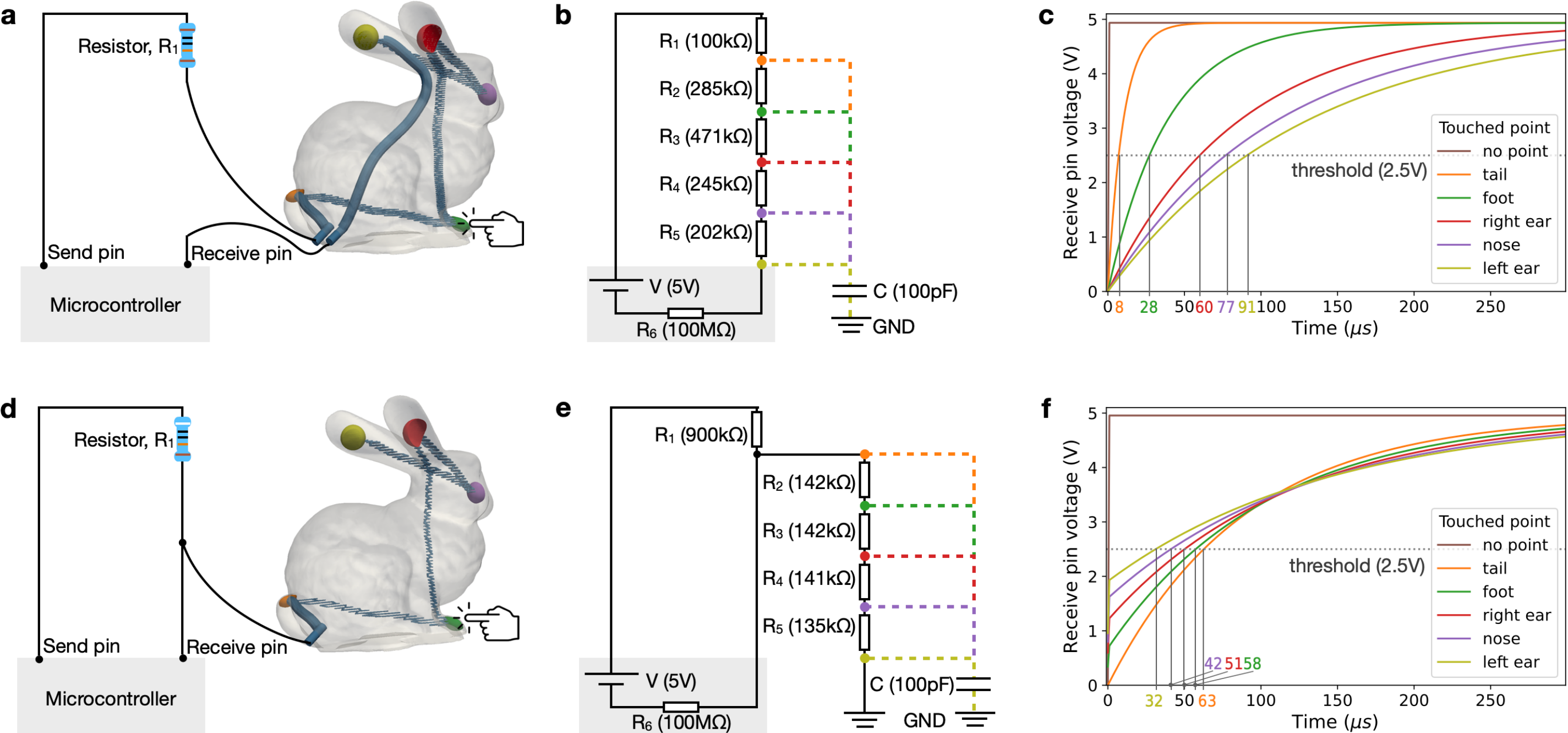}
    \caption{Capacitive sensing for the Stanford Bunny: (a) overall schematic on how the five touchpoints are connected via conductive traces (colored blue) and connected in series to a microcontroller's circuit using \textit{two} wires; (b) the circuit diagram corresponding to (a) with the representative resistance and capacitance measurements. Each colored dashed wire corresponds to a case when a point is touched (e.g., orange: tail, green: foot); and (c) the voltage change measured at the microcontroller's receive pin when a point is touched. 
    (d--f) correspond to (a--c) but the Stanford Bunny is connected in parallel to a microcontroller's circuit using \textit{one} wire.
    }
    \Description{Illustrations and graphs depicting the overall setup of connecting freeform interface to a microcontroller for a single-wire and double-wire configuration. Results of an electronic circuit with a microcontroller and sensors, demonstrate the unique RC delay time responses for each touchpoint.}
    \label{fig:principle}
\end{figure*}

To this end, our primary contribution is a computational design pipeline that leverages multi-material printing to embed multiple capacitive touchpoints into any 3D model that has a closed mesh without self-intersection (\autoref{fig:teaser}).
Our approach focuses on abiding by the given geometric constraint of a 3D model as opposed to modifying it.
After users select touchpoints and wiring connection point(s) on the model's surface, our pipeline employs a graph-based pathfinding algorithm to serially connect the touchpoints (\autoref{fig:teaser}b) and then uses the resulting path to generate conductive traces (i.e., 3D printed resistors) between each pair of touchpoints through a serpentine trace space-filling algorithm (\autoref{fig:teaser}c).
These conductive traces are optimized to achieve electrical resistance across the user-defined touchpoints to exploit a phenomenon called resistor-capacitor (RC) delays. 
By creating unique RC time delays for all touchpoints, each touchpoint can be capacitively sensed using only a \textbf{single-wire} or \textbf{double-wire} connection (\autoref{fig:principle}).

Achieving interactivity with a \singlewire\ is the extreme case of minimal instrumentation. Our secondary contribution is a thorough investigation of how to achieve this extrema and its mathematical and computational boundaries.
This approach enables interactivity in 3D printed objects for an extensive range of 3D geometry while even improving overall sensing reliability (cf. \autoref{sec8:robustness}).
Prior works have also leveraged multi-material 3D printing with conductive materials to create interactive objects. However, these objects either still require significant instrumentation (e.g., $n$ wires linked to a microcontroller to enable $n$ touchpoints)~\cite{schmitz2015capricate, palma2024capacitive, schmitz2019trilaterate, pourjafarian2019multi}.
Our work highlights how we can fabricate interactive objects irrespective of their complex geometry with minimal instrumentation.

We demonstrate the scalability, computational performance, robustness, accuracy, and applicability of our approach with corresponding technical evaluations. 
For scalability, our approach can embed 20 touchpoints into a 3D object using a \singlewire\ connection when there is at least \si{40\milli\meter} of distance between each pair of touchpoints. This distance can be further reduced to  \si{12\milli\meter} through parameter adjustments. Our sensing evaluation highlights real-time recognition of 93.35\% mean accuracy for the \singlewire\ connection and 89.49\% mean accuracy for the \doublewire\ connection by testing with 8 different objects. 
The higher accuracy of the \singlewire\ connection is further validated by our robustness evaluation. 
Also, through this robustness evaluation, we further discuss how to better improve the recognition accuracy for both \singlewire\ and \doublewire\ connections.
Our six applications---Stanford Bunny, MIDI Drumpads, Hilbert Curve, Chinese Character (Power), Chinese Lion, and Globe---demonstrate our method's flexibility in supporting a range of geometries with varying complexity (\autoref{fig:fab-objects}).

The source code of our computational design pipeline and supplemental materials can be found at \url{https://github.com/d-rep-lab/3dp-singlewire-sensing}. A video demonstration of our sensing technique can be found in the supplemental materials.

\section{Related Work}
\label{sec:related}
Our work builds on prior research that demonstrates ways to computationally design and fabricate interactive objects and capacitive sensors with 3D printing.

\subsection{Interactive 3D Prints Using Electronics}
Electronic components (e.g., motors, LEDs) offer versatile functionalities and are the backbone of most interactive devices we encounter.
The most popular interactivity mechanism for modern 3D printed objects is embedding or attaching off-the-shelf electronic components to 3D printed objects~\cite{ballagas2018design}. 
However, integrating off-the-shelf electronic components into 3D prints can be challenging. An individual must design an object around these components, ensuring that they can be inserted and wired accordingly post-fabrication~\cite{swaminathan2020optistructures, savage2013sauron, ballagas2018design, groeger2016hotflex, peng2015layered, palma2024capacitive}.

Computationally designing the location of electronic components can reduce design labor as well as minimize instrumentation. For example, SurfCuit~\cite{Umetani2017surfcuit} and MorphSensor~\cite{zhu2020morphsensor} allow makers to computationally preview component placement (e.g., resistors and integrated circuits) on the exterior surface of an object and then manually connect them with conductive tape once the object is 3D printed. DefSense~\cite{bacher2016defsense} computationally designs channels so that wires and sensors can be embedded into a 3D print to enable deformation sensing. Similarly, ModElec~\cite{he2022modelec} further reduces manual labor of wiring by generating 3D-printable conductive traces with A* search algorithm~\cite{hart1968formal}. 

\begin{figure*}[t!]
    \centering
    \includegraphics[width=\linewidth]{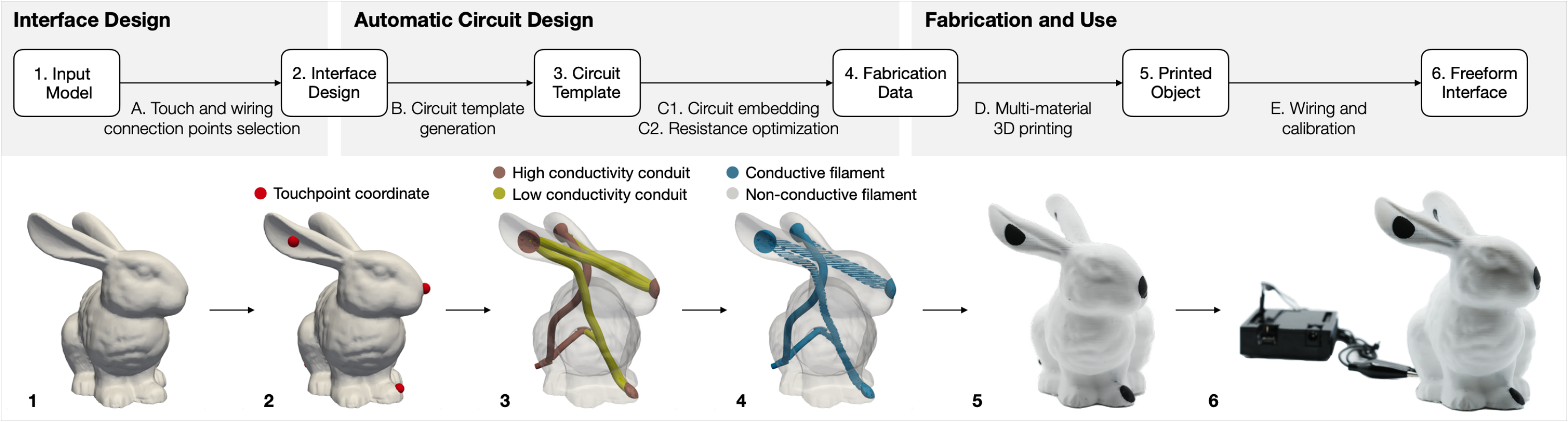}
    \caption{
    A computational design pipeline to create a freeform interface with embedded capacitive touchpoints.
    Rounded rectangles represent output data or physical objects. Arrows represent different processes. Each image at the bottom is the corresponding output data (i.e., rounded rectangles) from the computational pipeline.
    }
    \Description{
    Diagram illustrating the computational fabrication process in three stages. Stage 1 is the Interface Design where the designer will select touchpoint design to the input model. Stage 2 is Automatic Circuit design where circuit design is automatically routed throughout the bounding volume of the input model. Stage 3 is Fabrication and Use where the final freeform interface is attached to the microcontroller.
    }
    \label{fig:pipeline}
\end{figure*}

While these approaches help reduce design challenges, their methods still must account for the external physical electronic components. This reliance can influence the design of the object (e.g., prevent a small footprint) and still requires significant wiring and/or assembly, especially to integrate sensors. 
In contrast, our work aims to fabricate electronics as part of the 3D printing process, contributing to emerging research on 3D printable electronics~\cite{espalin20143d, FLOWERS2017156, Macdonald2014structural, goh20213d}. In our approach, conductive traces are automatically generated and 3D printed inside an object to act as resistors. This approach supports any 3D models that have a closed mesh without self-intersection, reduces the need for manual assembly, and minimizes the use of additional electronic components.

\subsection{3D Printed Capacitive Sensors}
\label{sec2:capacitive-sensors}
Capacitive sensing is a popular technique to capture touch input on devices by capacitively coupling the human body to a conductive material (e.g., an electrode or wire). We refer readers to \citet{grosse2017findingcommon}'s survey highlighting how capacitive sensing has been used in various human-computer interaction (HCI) contexts. Embedding conductive materials---including conductive filament---into 3D printed objects can enable capacitive sensing. These 3D printed objects generally fall under two categories: they are either designed with a conductive bottom surface that can be sensed on touchscreen devices~\cite{Schmitz2017flexibles, schmitz2021itsy} or with conductive regions that can be wired to an external microcontroller~\cite{schmitz2015capricate, bae2023computational, burstyn2015printput, palma2024capacitive}.

Our approach aligns with the second category.
Several prior works~\cite{schmitz2015capricate, burstyn2015printput,schmitz2019trilaterate, bae2023computational, palma2024capacitive, alalawi2023mechsense, takada2016mono, Kato2020CAPath,Ikematsu2018Ohmic}
demonstrate techniques to generate electrical traces within a 3D model. The resulting 3D printed objects have multiple touchpoints for sensing once they are connected to a microcontroller. However, the majority of these approaches~\cite{schmitz2015capricate, burstyn2015printput, schmitz2019trilaterate, palma2024capacitive, takada2016mono, Kato2020CAPath,Ikematsu2018Ohmic} still require significant instrumentation (e.g., $n$ wires connected to a microcontroller to sense $n$ touchpoints). 
In contrast, our work focuses on \textit{minimal instrumentation}, requiring only either a single-wire or double-wire connection(s) (\autoref{fig:principle}).
Our previous work~\cite{bae2023computational} also explored how to reduce instrumentation but is limited to only network-like geometry (i.e., spheres and cylinders)~\cite{rossignac2005shape}.
This severe constraint cannot generalize to arbitrary 3D forms. 
Our current approach lifts this constraint by supporting any closed, non-self-intersecting mesh, regardless of geometric or topological complexity. 
This capability expands the design space to fabricate complex geometric models with intrinsic interactivity, eliminating the need for post-processing.
Furthermore, we deepen the technical foundation of this minimal instrumentation approach by systematically optimizing the circuit design needed to enable accurate \singlewire{} sensing of multiple touchpoints (cf. \autoref{sec:one-wire}).

\section{Principle of Capacitive Sensing with RC Delay}
\label{sec3:rc-delay}
Our computational design pipeline enables embedding multiple capacitive touchpoints within a 3D object such that all touchpoints can be sensed using only a \singlewire\ or \doublewire\ connection (\autoref{fig:principle}). We provide a short introduction to capacitive sensing using RC delay, which is the key principle underlying our approach. 

In a capacitive sensing circuit, when a user touches a conductive element (e.g., electrode), the user's body and the element become capacitively coupled~\cite{grosse2017findingcommon}. This coupling induces an \textit{RC delay} in the sensing circuit. RC delay is the time required to charge a capacitor in a circuit through a particular amount of electrical resistance. Increasing the resistance in a circuit will generally increase the amount of time needed to charge the capacitor, thereby creating a larger RC delay. If each conductive element in a circuit needs a different amount of time to charge when touched, we can infer what is being touched by measuring the time needed to reach a predefined voltage threshold (e.g., 2.5V) on a microcontroller. In our pipeline, the electrical resistance for each conductive touchpoint is optimized to achieve a different RC delay by varying the length of the conductive trace between each pair of touchpoints. 

\autoref{fig:principle} illustrates this sensing principle with the Stanford Bunny as our freeform interface. As shown in \autoref{fig:principle}c,  \SI{0}{\micro\second} indicates the baseline in which no touchpoints are touched. Touching the bunny's tail requires \SI{8}{\micro\second} to reach the voltage threshold, while its foot requires \SI{28}{\micro\second}. 
Using this approach, we can detect multiple capacitive touchpoints by connecting the 3D printed object to a microcontroller with either \textit{two} (\autoref{fig:principle}a--c) or \textit{one} wire (\autoref{fig:principle}d--f). 
A \singlewire\ connection results in a parallel circuit, while the \doublewire\ connection results in a series circuit. The difference between these two circuit configurations results in different possible ranges of RC delays (\autoref{fig:principle}c vs. \autoref{fig:principle}f). 

\section{Computational Design Pipeline Overview}
\label{sec3:pipeline-overview}
The main objective of our computational pipeline is to embed multiple capacitive touchpoints into a freeform model. Our pipeline can work for any 3D models that have a closed mesh without self-intersection.
To achieve this goal, our computational pipeline (\autoref{fig:pipeline}) is divided into three stages: interface design (\autoref{sec3:interface-design}), automatic circuit design (\autoref{sec3:auto-circuit-design}), and fabrication and use (\autoref{sec3:physical-assembly}).

\paragraph{Interface Design.} In this first stage, the designer prepares a freeform model by either 3D modeling with a CAD software or uploading an existing 3D model.
Afterward, the designer (1) selects where the touchpoints will be on the 3D geometry's surface and (2) chooses the connection points (i.e., one or two) that will connect the 3D printed object to a microcontroller (see \autoref{fig:principle}a,d).

\paragraph{Automatic Circuit Design.}
After selecting the touchpoints and wiring connection point(s) on the freeform model, the second stage uses our computational algorithms to generate the necessary geometry to route conductive traces throughout the freeform model.
This step consists of two stages.
First, our graph-based pathfinding algorithm computes a path to serially connect all the touch and wiring points. 
Next, to produce sufficient resistance between each touchpoint for the RC delay, our space-filling algorithm draws long, thin conductive traces within the path. 

\paragraph{Fabrication and Use.} At the final stage, the designer uses the fabrication data from the second stage to print the model.
The 3D printed object is connected to a microcontroller with either one or two wires.
After calibrating all touchpoints to sense touches, the freeform interface is ready for use.

In the following sections, we use the Stanford Bunny with a \doublewire\ connection (\autoref{fig:principle}a) to illustrate the computational design pipeline. 
The \doublewire\ connection serves as the foundation to understand how we can implement a \singlewire\ design (\autoref{fig:principle}d).

\begin{figure}[t]
    \centering
    \includegraphics[width=\linewidth]{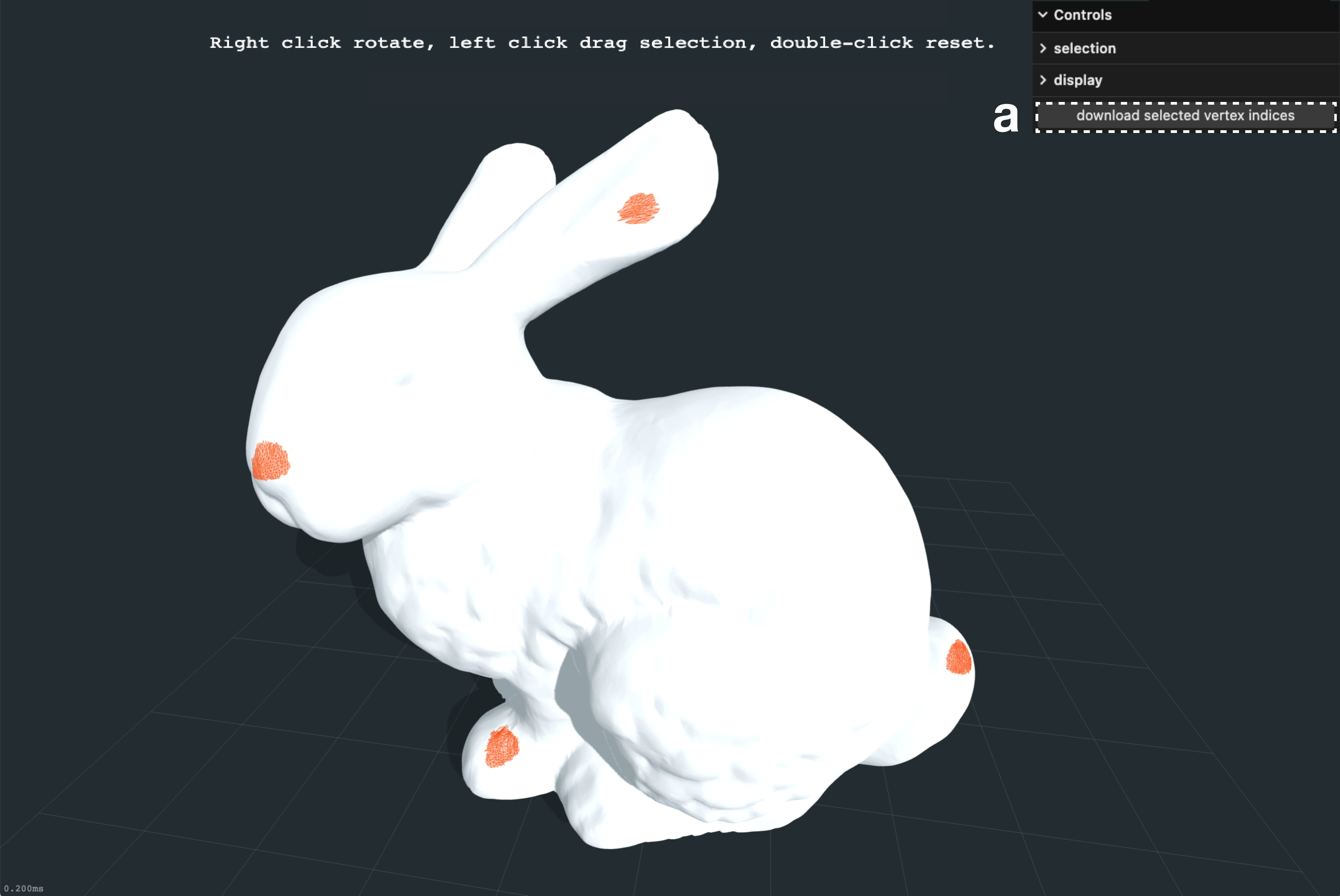}
    \caption{3D model of the Stanford Bunny with five selected touchpoints (foot, nose, left ear, right ear, and tail) and two wiring connection points. The right ear and the two wiring points are hidden from the viewpoint. The orange meshes indicate the user's lassoed selections. The dashed lines in (a) show where users can download the selected mesh coordinates.}
    \label{fig:web-ui}
    \Description{A graphical rendition of the Stanford Bunny on our web interface. The Stanford Bunny has five selected touchpoints (foot, nose, left ear, right ear, and tail) and two wiring connection points. All points are indicated with orange mesh. A drop-down menu is outlined with dashed lines indicating where users can download the selected mesh coordinates.}
\end{figure}

\section{Interface Design}
\label{sec3:interface-design}
The designer manually designs a freeform model or uses an existing model, and selects the coordinates of the touchpoints and wiring connection points on the 3D model's surface.
Our pipeline allows a designer to use any software of their choice (e.g., Fusion360, Rhino) to generate the coordinates.
We also provide a web-based user interface (UI) (\autoref{fig:web-ui}) to help facilitate this step. To select the touchpoints and wiring connection points, the web-based UI requires the following steps.

\paragraph{Upload STL File.} The designer uploads the STL file of a freeform model to the web-based UI. The uploaded file is then rendered as a 3D model for the designer to pan, rotate, and view.

\paragraph{Select Touchpoints and Wiring Connection Points.} After viewing the model, the designer can freely lasso different areas on the model's surface to indicate where the touchpoints and wiring connection points would be placed. \autoref{fig:web-ui} shows an example where the user has converted the Stanford Bunny's nose, foot, ears, and tail as touchpoints.

\paragraph{Export Coordinates.} Once the point selection has been finalized, the designer can export the points' coordinates (\autoref{fig:web-ui}a). The centroids of these coordinates will be used to generate touchpoints and wiring connection points on the surface of the freeform model.

\section{Automatic Circuit Design}
\label{sec3:auto-circuit-design}
The objective of the automatic circuit design stage is to generate the appropriate internal circuit design that will be embedded in the freeform model. This stage is fully automatic and does not require any active involvement from the interface designer.

\begin{figure}[!t]
    \centering
    \includegraphics[width=\linewidth]{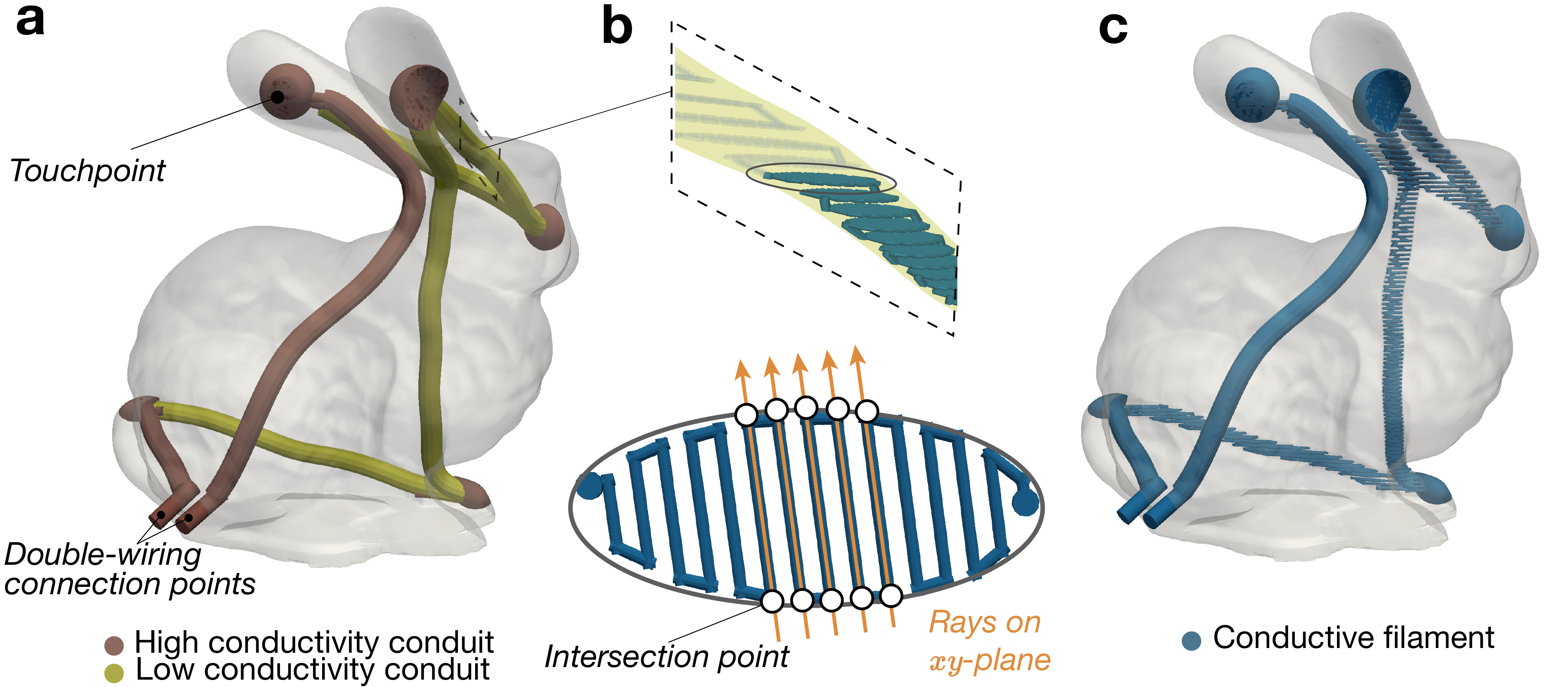}
    \caption{Automatic circuit design. (a) Our automatic circuit design first generates a circuit template, which outlines how we will embed the touchpoints and conductive traces inside a freeform interface.
    (b) During circuit embedding, serpentine trace patterns are generated inside low conductivity \pipes\ by using a space-filling algorithm. (c) The output is fabrication data (STL files), which will be used for multi-material printing.}
    \Description{Substages of our automatic circuit design process. The Stanford Bunny has two types of conduits routed throughout its bounding volume. A maroon color depicts the high conductivity conduit. A yellow color depicts the low conductivity conduit. A serpentine trace pattern shows how it is generated by casting rays on the xy-plane and finding its intersection point. }
    \label{fig:components}
\end{figure}

The automatic circuit design is broken into two sub-stages. The first step is to generate a circuit template (\autoref{sec3:circuit-template}). A circuit template specifies the geometry of points (touchpoints, wiring points) and \pipes\ (\autoref{fig:components}a).
This information is used to embed the internal circuit design into the freeform model in the subsequent step. Next, in the circuit embedding step (\autoref{sec:circuit-embedding}), our algorithm uses the information from the circuit template to draw the conductive traces for 3D printing using a space-filling algorithm (\autoref{fig:components}b,c).

For both substages, we use the following terminology: \textit{\points}\ referring to the touchpoints and the wiring connection point(s); \textit{\pipes}\ as the generated pipes within the freeform models' volume.

\begin{figure*}[t]
    \centering
    \includegraphics[width=\linewidth]{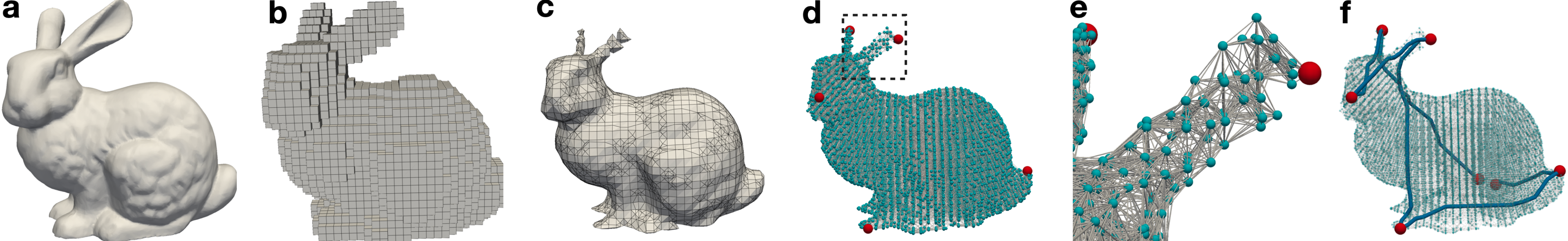}
    \caption{Representations involved in pathfinding: (a)  input 3D model for pathfinding; (b) voxel representation of the input model; (c) voxel representation after trimming voxels close to the model's surface; (d) graph representation of the trimmed voxel representation, where a vertex closest to each touchpoint is colored red; (e) a close-up look of the graph representation; and (f) identified paths using Dijkstra's pathfinding algorithm.
    Except for (e), all figures share the same camera position and angle.
    For presentation purposes, the voxel and graph representations have lower resolution (i.e., smaller numbers of voxels and vertices) than implemented.}
    \Description{The five representations used in our pathfinding algorithm. The first representation is the input 3D model. Then we transform the input model into a voxel representation. The third model is a trimmed version of the voxel representation. The fourth model is the graph representation of the model. The fifth representation is the graph representation with a found path.}
    \label{fig:pathfinding}
\end{figure*}

\subsection{Circuit Template Generation}
\label{sec3:circuit-template}
Generating a circuit template requires two sets of information.

\paragraph{Preparing Point Geometry.}
The first step of the circuit template generation is to prepare the geometry of the \points\ by using the coordinates of touchpoints and wiring connection points (cf. \autoref{sec3:interface-design}). 

\paragraph{Touchpoints.} We ensure that the geometry of touch\points\ is bounded within the freeform model's volume with two steps. The first step involves generating a 3D geometry at the centroid of each touchpoint's coordinate. We default to a sphere with a \si{12\milli\meter} diameter, but other sizes and types of 3D geometries can also be considered depending on the designer's needs.
We then volumetrically clip the sphere based on its intersection with the freeform model's surface. 

\paragraph{Wiring Connection Points.} For the wiring connection point(s), we generate a cylinder that intentionally extrudes beyond the model's surface (\autoref{fig:components}a). A cylindrical design allows external wires (e.g., alligator clips) to easily connect to the freeform model at the fabrication and use stage. 
To achieve this cylindrical design, we first compute both the centroid and normal of a polygon at the specified coordinates of the wiring connection point(s). 
We then generate a cylinder with a \si{4\milli\meter} diameter and that is \si{10\milli\meter} long. These dimensions are arbitrary and sufficient for an alligator clip to grip onto. We use the centroid as the cylinder's center and the normal as the cylinder's axis. 
The cylinder's height, diameter, and axis direction can also be manually specified.

\paragraph{Routing Conduits with Pathfinding Algorithm.}
Next, we generate \pipes\ (i.e., 3D pipes). The \pipes\ serve two purposes. 
First, they will connect the \points\ within the 3D model's bounding volume. 
Second, they will house the 3D printed conductive traces. 

We first need to consider how to connect \points. 
Although the connection of \points\ can be either in series or parallel, we focus only on a series connection.
A series connection is simpler to design for an RC circuit as well as controlling the RC delay. 
In addition, based on our assumption that the freeform model has a closed mesh without self-intersection, we expect the model to have enough volume to construct a series connection within the model.

To make a series connection, we need to decide the order of the \points. 
In the case of the \doublewire\ connection, the first and last \points\ are the wiring connection points. For the \singlewire\ connection, the first point corresponds to the wiring connection point. 
The designer can manually specify which \points\ are used as the wiring connection point(s).
By default, the remaining \points\ (i.e., touchpoints) are connected in the selected order during the interface design stage.
For example, the point order for \autoref{fig:components}a is the following: first wiring connection point, tail, foot, right ear, nose, left ear, and second wiring connection point.

We then route the \pipes\ to connect the \points\ in the specified order with our graph-based pathfinding algorithm.
Our pathfinding algorithm consists of four steps: (1) voxelize the freeform model, (2) trim the voxels that are close to the freeform model's surface, (3) construct a weighted neighbor graph with the remaining voxels, and (4) find the shortest path between each point to connect all \points. 
These steps are visually summarized in \autoref{fig:pathfinding}.
The shortest path is used to ensure there is sufficient space to generate other conduits with the remaining bounding volume.
To help distinguish from the terminology used for the circuit template design (i.e., \points, \pipes), we use \textit{graph}, \textit{vertices}, and \textit{edges} as specific terminology for our pathfinding algorithm.

\begin{enumerate}
    \item \textit{Voxelize the freeform model (\autoref{fig:pathfinding}b).}
    Voxelization is necessary to prepare a graph that will route the \pipes\ inside the freeform model's bounding volume.
    We generate a voxel representation of the model using a specified voxel size. By default, our pipeline uses 0.5\% of the maximum dimension of any side of the model’s bounding box.
    
    \item \textit{Trim the voxels close to the model surface (\autoref{fig:pathfinding}c).}
    If a \pipe\ is placed too close to the model surface, it can introduce parasitic capacitance (i.e., unintentional capacitance) during use. 
    Parasitic capacitance is non-ideal as it can influence the overall sensing capability.
    Thus, we trim the voxels that are too close to the model surface (by default, \si{3\milli\meter} from the surface). 
    
    \item \textit{Construct a weighted neighbor graph (\autoref{fig:pathfinding}d,e).}
    We generate a weighted neighbor graph from the trimmed voxel representation.
    We first compute the distance between each voxel and then construct a $k$-nearest neighbor graph based on the distances ($k=10$ by default).
    This process generates a graph consisting of vertices corresponding to the voxels, the edges of neighbor relationships, and the edge weights corresponding to the distances.
    
    \item \textit{Find the shortest paths (\autoref{fig:pathfinding}f).}
    We can find the shortest path between a pair of vertices using a pathfinding algorithm such as Dijkstra's algorithm or A*~\cite{hart1968formal}. 
    By default, we employ Dijkstra's algorithm, but our implementation is flexible to switch to other pathfinding algorithms. 
    We iteratively perform this pathfinding step to find the route that connects all \points\ in series. 
    In parallel, our algorithm aims to avoid overlapping \pipes\ with each other. Since \pipes\ will house the conductive traces, an overlap can change the circuit design.
    After each iteration, we assign a large penalty for the edge weights (\SI{300}{\milli\meter}) that have already been used or are too close to the found path.
    Although the shortest path is generally preferable to ensure sufficient space, the path between two \points\ may be too short to generate sufficiently large resistance in the circuit embedding step (\autoref{sec:circuit-embedding}).
    To resolve such a case, we provide two options. 
    The first option is to randomly permute the connection order of touchpoints until all paths become longer than the designer-specified lengths.
    The second option is to run our shortest path finding algorithm multiple times.
    Due to the penalty added in the edge weights, the algorithm can make a path gradually longer.
\end{enumerate}

After the route of the \pipes\ is finalized, the \pipes\ are then rendered as 3D pipes (\si{{5}\milli\meter} diameter by default). We chose \si{{5}\milli\meter} as our default to provide enough space to generate the serpentine trace patterns with our nozzle's extrusion width (\si{{0.4}\milli\meter}).

\subsection{Circuit Embedding}
\label{sec:circuit-embedding}
We use the circuit template to generate the freeform model's internal circuit design.
3D printing the circuit requires a combination of conductive and non-conductive materials.
As mentioned, \points\ are the touchpoints and wiring connection points are filled with a large amount of conductive filament (100\% infill) by default. As a result, \points\ have high conductivity and negligible resistance. In contrast, the generated \pipes\ can have either \textit{high conductivity} or \textit{low conductivity} (\autoref{fig:components}a). The conductivity is determined based on the \pipe's role in the internal circuit.

Conduits that originate from the wiring connection point(s) have high conductivity (brown links in \autoref{fig:components}a).
These \pipes\ are meant to act as wires (i.e., negligible resistance). Similar to \points, the high conductivity \pipes\ will also be 3D printed with a 100\% infill with a conductive filament. In comparison, the \pipes\ between each pair of touchpoints have low conductivity (yellow links in \autoref{fig:components}a).
The low conductivity \pipes\ act as \textit{resistors}. To leverage RC delay for capacitive sensing, these low conductivity \pipes\ need to achieve high resistance within their limited volume.

We achieve high resistance by drawing a thin, long trace of the conductive filament using a serpentine trace pattern~\cite{soh2009comprehensive} inside the low conductivity \pipes\ (\autoref{fig:components}b). Due to the resistivity law, a thinner conductive trace will provide lower conductivity and higher resistance.
The thickness of the conductive traces varies when drawing the trace on the $xy$-plane versus along the $z$-direction. 
The thickness for the $xy$-plane can be close to the printer's nozzle extrusion width (e.g., \si{{0.8}\milli\meter}), while the thickness for $z$-direction should be at least twice the extrusion width (e.g., \si{{1.2}\milli\meter}) to ensure contact with the previously printed layer. The variance in thickness is to account for the printing resolution of common FDM printers.

To support a wide range of geometry, our pipeline needs to be able to handle curved \pipes. 
We designed a space-filling algorithm to draw serpentine trace patterns in these curved \pipes.
For a given $z$-coordinate along the $xy$-plane, we cast multiple rays that are parallel to each other. 
Each ray finds the intersection points with the \pipe\ and creates line segments by connecting the intersection points.
By alternatively connecting one line segment's endpoint and another line segment's start point, we can obtain a serpentine pattern for one layer (\autoref{fig:components}b). 
We repeat this process while gradually increasing (or decreasing) the $z$-coordinate, resulting in multiple layers of serpentine patterns. 
We then connect these serpentine patterns with a staircase pattern using vertical lines along the $z$-direction.
We also need to consider how much of a margin should be between each ray as well as between each layer. 
The margin must be larger than the printer's nozzle extrusion width. 
Based on the specifications of most FDM printers, we use 
\si{{1.2}\milli\meter} as the margin for both the ray and layer by default.
    
After generating all circuit components described above, the internal circuit design is output as STL files for multi-material 3D printing.

\subsection{Supporting Cases Using a Single-Wire Connection}
\label{sec:one-wire}
The serpentine pattern described in \autoref{sec:circuit-embedding} aims to achieve high resistance for the conductive traces in the low conductivity \pipes. 
Our analysis reveals that we can support the \doublewire\ connection as long as \textit{each} low conductivity \pipe\ has sufficiently high resistance (\autoref{sec:one-vs-two}).
However, to support a \singlewire\ connection, there are additional requirements: we need to carefully control the \textit{interplay of all} resistances in the circuit.
This additional requirement stems from how the \singlewire\ connection results in a parallel circuit, while the \doublewire\ connection results in a series circuit (see \autoref{fig:principle}).
To handle the differences in the overall circuit configuration, we first discuss the theoretical differences between the \doublewire\ and \singlewire\ connections.
We then introduce an optimization method to support multiple capacitive touchpoints with a \singlewire\ connection.  

\subsubsection{Single-Wire vs. Double-Wire Connections}
\label{sec:one-vs-two}

\paragraph{Double-Wire Connection.} 
When a freeform model is connected to the microcontroller with two wires like \autoref{fig:principle}a, we only need to ensure each low conductivity conduit has sufficiently large resistance (e.g., \si{50\kilo\ohm}, cf. \autoref{sec4:scalability}).
The large resistance allows a microcontroller to have enough buffer to capture the RC delay differences among the touchpoints.
We now discuss in detail the reasoning behind this simple requirement.

\begin{figure}[t]
    \centering
    \includegraphics[width=\linewidth]{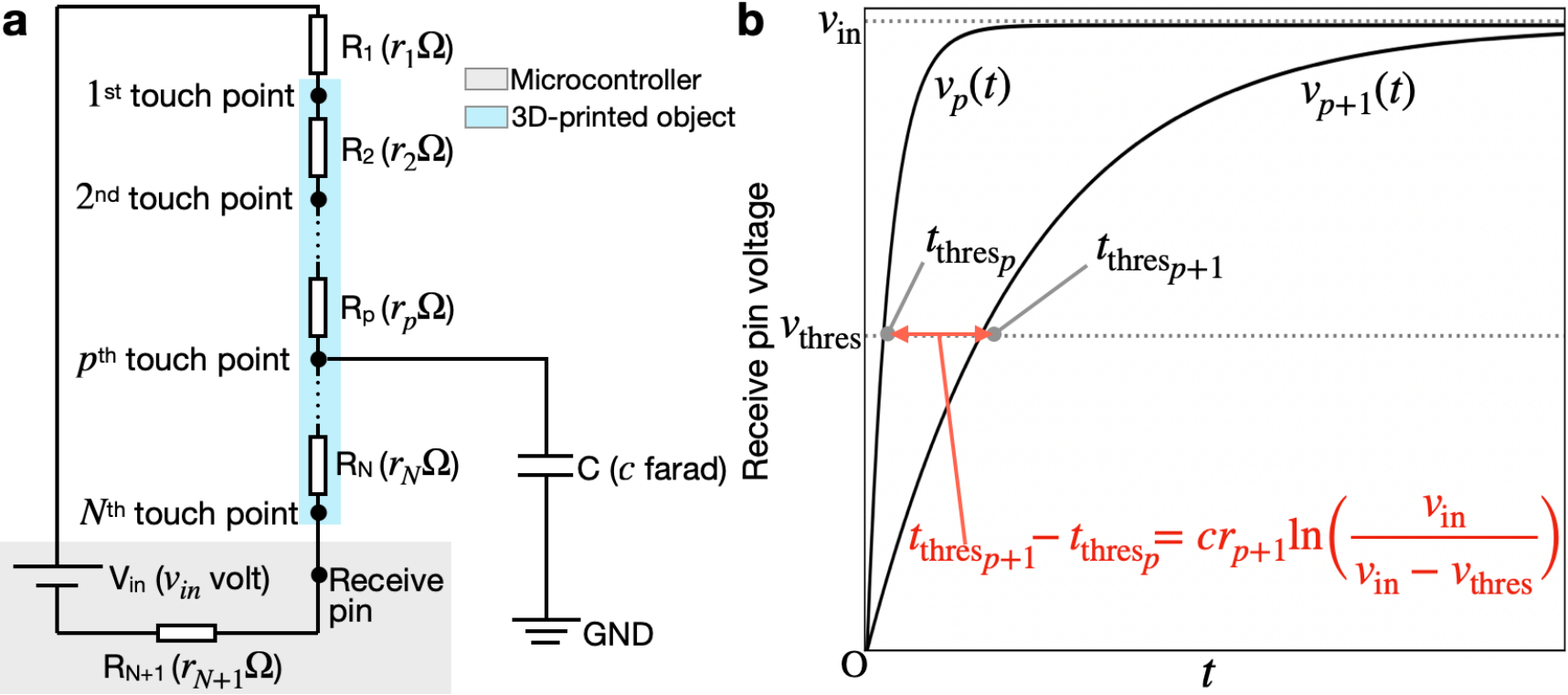}
    \caption{Analysis of a \doublewire\ connection. (a) shows the circuit schematic for the \doublewire\ connection, where the $p$th touchpoint is selected. (b) compares the voltage changes and time delays induced when $p$th and $(p+1)$th touchpoints are touched.}
    \Description{This image is a schematic representation of the double-wire connection, illustrating the overall circuit schematic and signal response. On the left (a), a diagram shows multiple touch points are connected serially along a resistive path connected to a microcontroller, with a ground reference.  On the right (b), a graph depicts the voltage change over time at the receive pin as a response to touch, highlighting the threshold voltage levels for different touch points.}
    \label{fig:double-wire}
\end{figure}

We assume the circuit shown in \autoref{fig:double-wire}a, where $R_1$ is a resistor connected to a microcontroller and $R_2, \cdots, R_\nTouchpoints$ ($\nTouchpoints$: the number of touchpoints) are the resistors embedded into the 3D printed object.
The microcontroller has a voltage source with $\VIn{}$ and a resistor between its voltage source and receive pin. 
For convenience, we denote this microcontroller's resistor as $R_{N+1}$.
Also, we denote $R_i$'s resistance as $r_i$ ($i = \{1, \cdots, N+1\}$).
In addition, the capacitor formed by touch has capacitance $c$.

When touching the $p$th touchpoint ($1 \leq p \leq \nTouchpoints)$, the voltage change measured at the receive pin can be written as:
\begin{equation}
\label{eq:voltage-transient-two-wires}
    v_p(t) = v_{\mathrm{in}} \left(1 - \exp \left(-\frac{t r_\mathrm{all}}{c \RTill{p} \RAfter{p}} \right)\right) \frac{r_{\nTouchpoints+1} \RTill{p}^2 \RAfter{p}^2}{r_\mathrm{all} \left(\sum_{i=0}^{p} r_i \RAfter{p} \right)^2}
\end{equation}
where $t$ is the time to charge a capacitor; $r_\mathrm{all} = \sum_{j=1}^{\nTouchpoints+1} r_j$; $\RTill{p} = \sum_{j=1}^{p} r_j$; and $\RAfter{p} = \sum_{j=p+1}^{\nTouchpoints+1} r_j$ (see \autoref{appendix-a:equation-derivation} for the derivation).
To induce a high-impedance state for the receive pin, $r_{\nTouchpoints+1}$ is usually very large (e.g., \si{100 \mega\Omega}). This value is pre-determined by the microcontroller's manufacturer.
When $r_1, \cdots, r_\nTouchpoints$ are relatively small (e.g., \si{100\kilo\Omega}), we can approximate \autoref{eq:voltage-transient-two-wires} as:
\begin{equation}
\label{eq:voltage-transient-two-wires-approx}
    v_p(t) \approx v_{\mathrm{in}} \left(1 - \exp \left({-\frac{t}{c \RTill{p} }} \right)\right)
\end{equation}

From \autoref{eq:voltage-transient-two-wires-approx}, we can approximate the time required to reach a microcontroller's logic threshold voltage, $\VThres{}$ as followed:
\begin{equation}
\label{eq:time-two-wires-approx}
    {t_\mathrm{thres}}_p \approx c \RTill{p} \ln \left(\frac{\VIn{}}{\VIn{} - \VThres{}} \right)
\end{equation}

\autoref{fig:double-wire}b summarizes the key theoretical relationships that we derived from the equations above.
Using \autoref{eq:time-two-wires-approx}, we can consider that the RC delay only depends on $\RTill{p}$, the cumulative sum of resistance values involved from the voltage source to a selected touchpoint. 
We further derive that ${t_\mathrm{thres}}_{p+1} - {t_\mathrm{thres}}_{p} = c r_{p+1} \ln ( \VIn{} / (\VIn{} - \VThres{}) )$.
This equation indicates that larger resistance for each of $r_1, \cdots, r_\nTouchpoints$, results in larger differences between ${t_\mathrm{thres}}_p$ and ${t_\mathrm{thres}}_{p+1}$.
Note: ${t_\mathrm{thres}}_{1} \leq {t_\mathrm{thres}}_2 \leq \cdots \leq {t_\mathrm{thres}}_\nTouchpoints$ because $c \geq 0$, $r_p \geq 0$, and $\ln(\VIn{}/ (\VIn{} - \VThres{})) \geq 0$. 

From the observations above, we can support the \doublewire\ connection by generating a serpentine trace pattern that is as long as possible within a conduit's volume (\autoref{sec:circuit-embedding}).
Thus, we only need to ensure that ${t_\mathrm{thres}}_{p+1} - {t_\mathrm{thres}}_{p}$ is large enough for a microcontroller to measure (e.g., 200 clock cycles of a microcontroller's CPU).
Note that when $\RTill{p}$ is extremely large, ${t_\mathrm{thres}}_p$ may be too large to provide reasonable latency for interactivity (e.g., when ${t_\mathrm{thres}}_p >$ \si{10\milli\second}).
However, the possibility of this situation is rare.
For example, a hypothetical scenario where ${t_\mathrm{thres}}_p >$ \si{10\milli\second} would require $\RTill{p} >$ \si{140\mega\Omega} when using the Arduino UNO R4 as the microcontroller ($\VIn{}{=}$\si{5\volt}, $\VThres{}{=}$\si{{2.5}\volt}) and $c=$\si{100\pico\farad} as a representative capacitance~\cite{esd2010fundamentals}.
Generating conductive traces where $\RTill{p} >$ \si{140\mega\Omega} inside a freeform model is almost infeasible.

\begin{figure}[t]
    \centering
    \includegraphics[width=\linewidth]{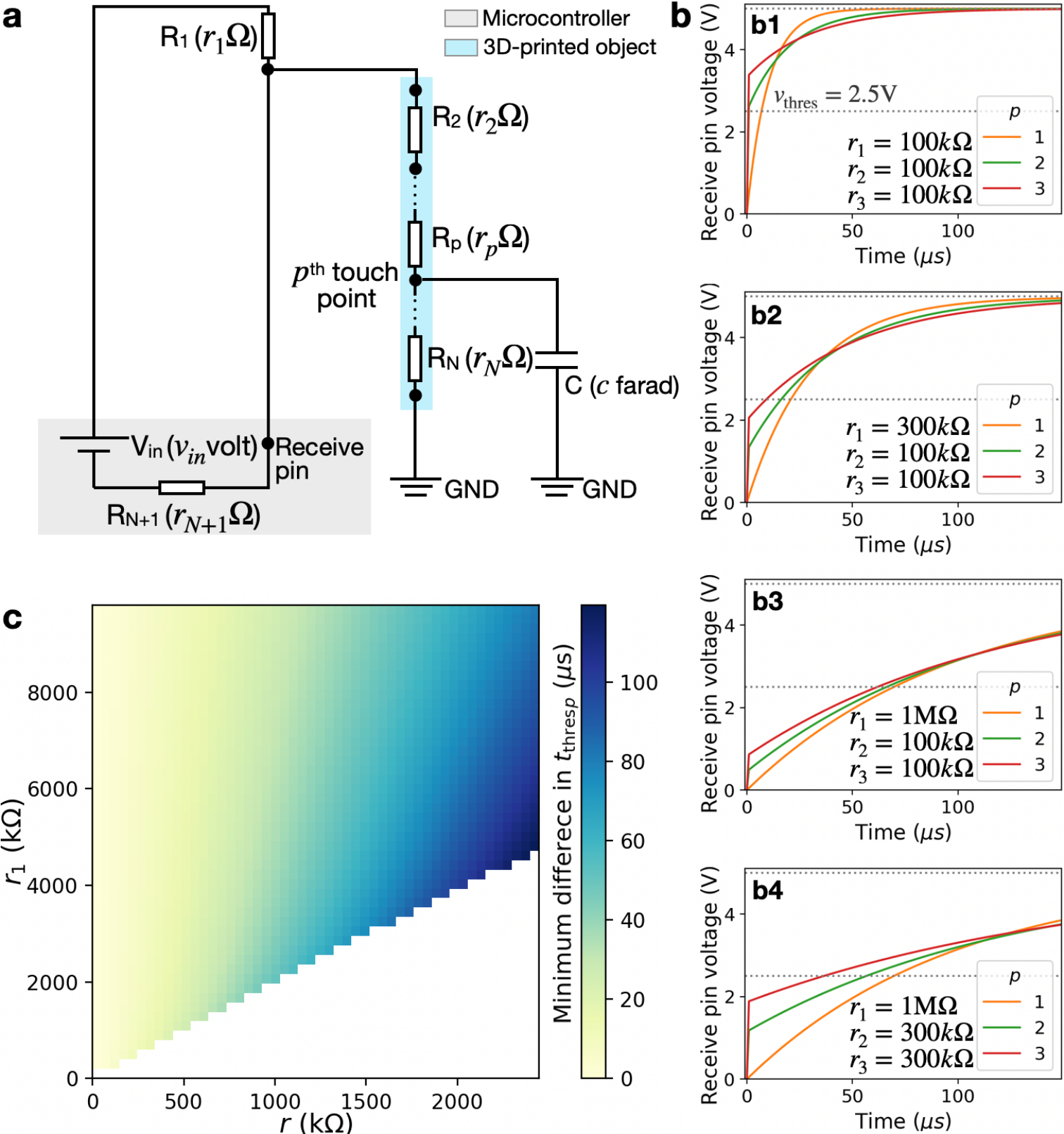}
    \caption{Analysis of a \singlewire\ connection. (a) shows the circuit schematic for the \singlewire\ connection where the $p$th touchpoint is selected. (b) compares the voltage changes across different variations of three resistance values ($r_1$, $r_2$, $r_3$). (c) shows the minimum differences among ${t_\mathrm{thres}}_1$, ${t_\mathrm{thres}}_2$, ${t_\mathrm{thres}}_3$ for the different variations of $r_1$ and $r$ (note: here $r_2 {=} r_3 {=} r$). A white grid cell indicates a violation of the hard constraint of $r_1$.
    For (b) and (c), we assume a case using the Arduino UNO R4 (i.e., $\VIn{} = 5\mathrm{V}$, $\VThres{} = 2.5\mathrm{V}$) and $c=100\mathrm{p}\mathrm{F}$.}
    \Description{This image is a schematic representation of the single-wire connection, illustrating the overall circuit schematic and it's voltage threshold responses. On the right is a column of four line plots illustrating voltage changes across different variations of three resistance values. Only one line plot has maximum distinct time differences for each voltage change. A grid cell in the bottom-left corner of the figure has a divergent color palette of yellow to blue. The lower right of the grid is colored white. These white grid cells indicate a violation of the hard constraint imposed on our capacitive sensing. }
    \label{fig:single-wire}
\end{figure}

\paragraph{Single-Wire Connection.}
For the \singlewire\ connection, we connect the freeform model to a microcontroller as shown in \autoref{fig:single-wire}a.
This connection makes a branch in the electrical path (i.e., forming a parallel circuit).
Consequently, for the \singlewire\ connection, we obtain the measured voltage change at the receive pin as:
\begin{equation}
\label{eq:voltage-transient-one-wire}
\resizebox{0.47\textwidth}{!}{$
    v_p(t) = v_{\mathrm{in}} \frac{1}{r_1 + r_{\nTouchpoints+1}} \left(r_{\nTouchpoints+1} - \frac{r_1 r_{\nTouchpoints+1}^2}{r_1 \RTill{p} + r_{\nTouchpoints+1} \RTill{p} - r_1^2} \exp \left(-\frac{t (r_1 + r_{\nTouchpoints+1})}{c (r_1 \RTill{p} + r_{\nTouchpoints+1} \RTill{p} - r_1^2)} \right)\right)
$}
\end{equation}
Similar to the \doublewire\ connection, we can approximate \autoref{eq:voltage-transient-one-wire} as:
\begin{equation}
\label{eq:voltage-transient-one-wire-approx}
    v_p(t) \approx v_{\mathrm{in}} \left(1 - \frac{r_1}{\RTill{p}} \exp \left({-\frac{t}{c \RTill{p} }} \right)\right)
\end{equation}
Then the approximate time required to reach a microcontroller's logic threshold voltage is:

\begin{equation}
\label{eq:time-one-wire-approx}
    {t_\mathrm{thres}}_p \approx c \RTill{p} \ln \left(\frac{r_1}{\RTill{p}} \frac{\VIn{}}{\VIn{} - \VThres{}} \right)
\end{equation}
We can observe that the difference between \autoref{eq:voltage-transient-one-wire-approx}--\ref{eq:time-one-wire-approx} and \autoref{eq:voltage-transient-two-wires-approx}--\ref{eq:time-two-wires-approx} is the coefficient $r_1/ \RTill{p}$ (or $\RTill{p}/ r_\mathrm{r_1}$).
This coefficient introduces greater complexity for the \singlewire\ connection compared to the \doublewire\ connection. 

The most critical difference from the \doublewire\ connection is the voltage measured at the receive pin at $t{=}0$: $v_p(0) = \VIn{}(1 - r_1 / \RTill{p})$.
This indicates $v_p(0)$ changes depending on the relationships between $r_1$ and $\RTill{p}$ (note: $\RTill{p} = r_1 + \cdots + r_p$).
Additionally, the coefficient, $r_1/ \RTill{p}$, influences the slope of the exponential function in \autoref{eq:voltage-transient-one-wire-approx}.
These facts suggest that we need to carefully select $r_1$ and $\{r_2, \cdots, r_n\}$ to support the \singlewire\ connection. This selection requires two considerations.

First, we have a hard constraint for $r_1$.
To sense a selected touchpoint, we must avoid where $v_p(0) \geq \VThres{}$ (i.e., $r_1/\RTill{p} \leq 1 - \VThres{} / \VIn{}$).
For example, \autoref{fig:single-wire}-b1 reflects this violation, and a microcontroller would not be able to detect if a touchpoint has been selected for two of the points (i.e., $p=1$ and $p=2$). 
Thus, to ensure all touchpoints can be sensed, we must satisfy $r_1/\RTill{p} > 1 - \VThres{} / \VIn{}$ for all touchpoints.
If using the Arduino UNO R4, $1 - \VThres{} / \VIn{} = 0.5$; thus, the hard constraint corresponds to $r_1 > r_2 + \cdots + r_\nTouchpoints$. 
This indicates that $r_1$ must be greater than the cumulative sum of the resistance inside the freeform model.

Second, to maximize the difference in ${t_\mathrm{thres}}_p$ for each touchpoint (e.g., $p$th vs. $(p+1)$th touchpoint), we need to resolve the complex relationships among $r_1$ and $r_2, \cdots, r_\nTouchpoints$. 
\autoref{fig:single-wire}b shows four variations that use different resistance values for the same example. 
Among the four variations, \autoref{fig:single-wire}-b4 achieves the maximum difference in ${t_\mathrm{thres}}_p$ for each touchpoint. 
To find the configuration with the largest difference in ${t_\mathrm{thres}}_p$ for each touchpoint, we introduce a heuristic resistance optimization for the \singlewire\ connection.

\subsubsection{Resistance Optimization}
\label{sec:one-wire-optimization}
Our heuristic resistance optimization for the \singlewire\ connection consists of two steps: (1) identify the appropriate resistance values and (2) adjust the geometry of the serpentine trace patterns. This adjustment will take place during the circuit embedding step described in \autoref{sec:circuit-embedding}.

\paragraph{Identify Appropriate Resistance Values.} 
The objective of the first step is to optimize $r_1$ and $r_2, \cdots, r_\nTouchpoints$. The goal is to \textit{maximize the minimum difference among the time delays}. This goal ensures the time delay difference for each touchpoint is large enough for a microcontroller to measure.
To heuristically achieve this goal, we perform a grid search utilizing a circuit simulator, specifically, Lcapy~\cite{hayes2022lcapy}.
To make the search space reasonably small, we consider a case where all resistance values within the freeform model have the same value, i.e., $r_2 = \cdots = r_\nTouchpoints = r$. 
We then only have two parameters, $r_1$ and $r$, to search for a given $\nTouchpoints$ (i.e., the number of touchpoints), $\VIn{}$, and $\VThres{}$. 
We satisfy $r_2 = \cdots = r_\nTouchpoints = r$  when adjusting the serpentine trace pattern's geometry in the subsequent step. 

We must first specify the search range and step for each $r_1$ and $r$. 
For $r_1$, we set the search range as [\si{200\kilo\Omega}, \si{10\mega\Omega}] and the step increment as \si{200\kilo\Omega} by default.
These values were chosen in balance of computational performance and optimization quality.
For $r$, we first identify the highest resistance each \pipe\ can achieve with the serpentine trace pattern. 
Among these values, we select the lowest value as the upper limit for $r$. 
The step increment is set as \si{50\kilo\Omega}.
For each grid cell (e.g., $r_1=$ \si{1\mega\Omega} and $r=$ \si{100\kilo\Omega}), we use a circuit simulator to generate $\nTouchpoints$ circuits, each of which corresponds to a case where the $p$th point is touched ($1 \leq p \leq \nTouchpoints$). 
Among the $\nTouchpoints$ different ${t_\mathrm{thres}}_p$ values, we select the pair that has the minimum difference.
Each non-white cell in \autoref{fig:single-wire}c illustrates this minimum difference.
The optimal result in \autoref{fig:single-wire}c is the cell with the largest $r$ and smallest $r_1$ (i.e., $r = \si{2500\kilo\ohm}$, $r_1 = \si{4200\kilo\ohm}$). 
However, when $r_1$ is near the boundary of the hard constraint, $v_p(0)$ for the $\nTouchpoints$th touchpoint is also close to $\VThres{}$ (e.g., \autoref{fig:single-wire}-b2).
Cases when $v_N(0)$ is close to $\VThres{}$ are problematic: they can introduce violations where $r$ may be larger than expected. This violation can be further exacerbated when 3D printing the conductive traces with poor precision.
To account for general fault tolerance in 3D printers, we select a pair of $r_1$ and $r$ that achieves a close-to-optimal result while satisfying the condition that $v_p(0) \leq 0.9 \VThres{}$.  

\paragraph{Adjust the Serpentine Trace Patterns.}
After optimizing $r_1$ and $r$, we apply these values to the circuit design.
$r_1$ is the resistance of an outside resistor connected to a microcontroller, and thus, it can be easily adjusted by hand.
In contrast, $r$ is the resistance value for each low conductivity \pipe, and we can achieve $r$ by adjusting the serpentine trace patterns.
As discussed in \autoref{sec:circuit-embedding}, the circuit embedding step aims to generate the longest conductive trace (i.e., largest resistance) by filling a serpentine trace pattern with a given small margin (by default, \si{{1.2}\milli\meter}). 
We can find the serpentine trace pattern that achieves $r_2 = \cdots = r_n = r$ by gradually increasing both the margin between each ray and the margin between each layer.

\section{Fabrication and Use}
\label{sec3:physical-assembly}

\paragraph{Multi-Material Printing.}
\label{sec3:multi-material}
After automatically designing the circuit, the computational pipeline outputs the fabrication data as four STL files: the original 3D model, the conductive traces, the touchpoints and wiring connection point(s), and the \pipes\ to encase the conductive traces. We use all four files for fabrication. In \autoref{appendix-b:implementation}, we discuss in more detail our 3D print settings and the filaments that we use. See our Github repository\footnote{\url{https://github.com/d-rep-lab/3dp-singlewire-sensing}} for the STL files of our freeform interfaces.

\paragraph{Wiring and Calibration.}
\label{sec3:wiring}
After 3D printing, we connect the printed object to a microcontroller. \sloppy{The schematic differs whether the freeform interface uses a \singlewire\ or \doublewire\ connection.}
\autoref{fig:principle}a and \autoref{fig:principle}d represents the schematic diagram for the \doublewire\ and \singlewire\ connection, respectively.
Lastly, using an existing signal processing library~\cite{bae2023computational}, we calibrate the RC delay corresponding to each touchpoint by manually touching each touchpoint for five seconds and observing the time required to reach a microcontroller's logic threshold voltage.
Calibration is necessary as each individual and external factors (e.g., clothing, temperature) may generate a different capacitance~\cite{grosse2017findingcommon}. A video demonstration of our sensing technique can be found in the supplemental video.
\begin{figure*}[htbp]
    \centering
    \includegraphics[width=0.90\textwidth]{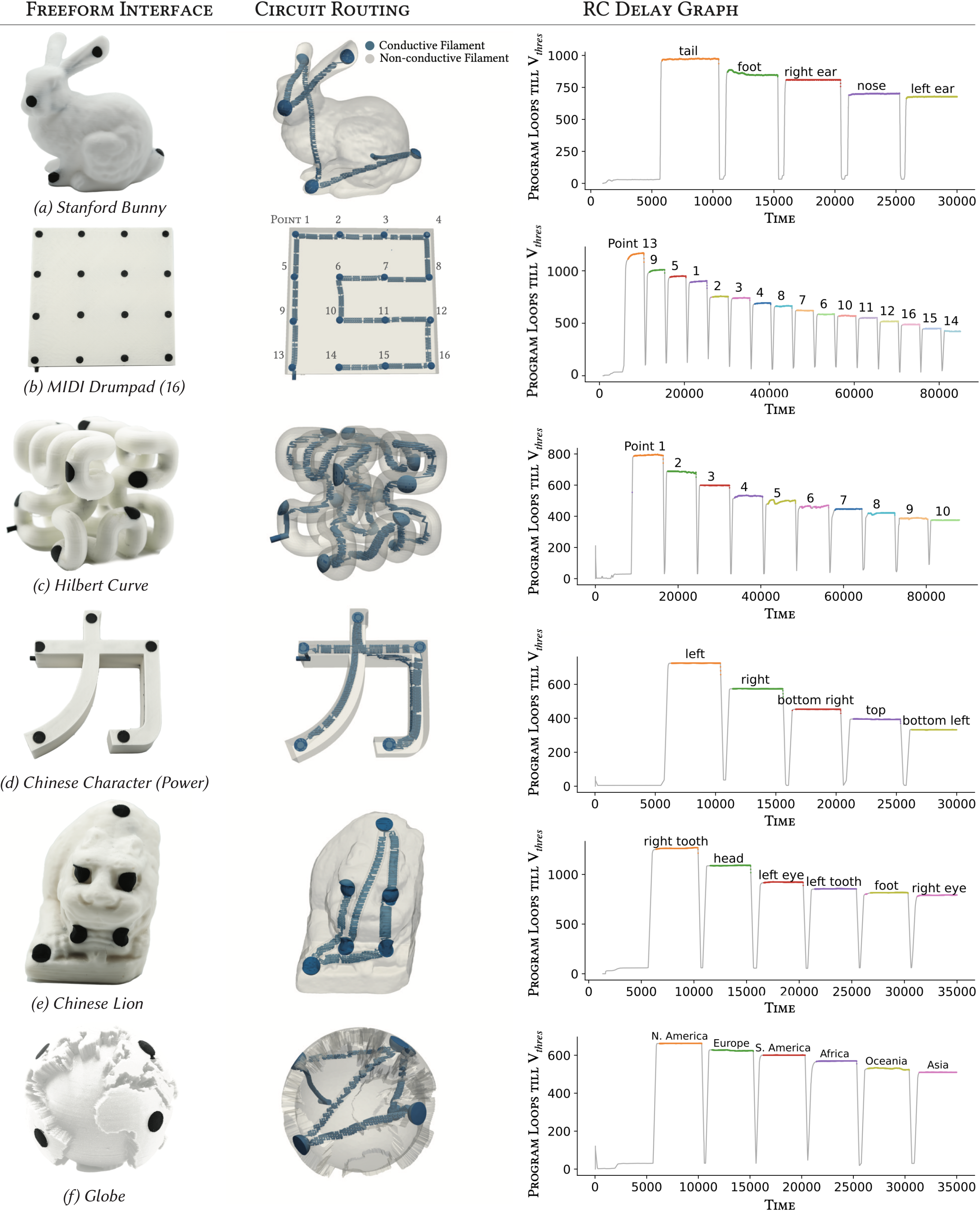}
    \caption{Examples of freeform interfaces with a \singlewire\ connection designed with our computational design pipeline. Each row shows the 3D printed object, its internal circuit design, and its corresponding RC delay graph. The RC delay graph is made by touching each touchpoint one after another. The $x$-axis corresponds to the time elapsed since using the freeform interface. The $y$-axis corresponds to the number of program loops that elapsed until reaching the microcontroller's voltage threshold. 
    The RC delay graph is colored based on each touchpoint's unique RC delay (e.g., gray: no touch, orange: first touchpoint).}
    \Description{ A table with three columns and 6 rows. Each row illustrates an example of a freeform interface with a single-wire connection (Stanford Bunny, MIDI Drumpad (16), Hilbert Curve, Chinese Character (Power), Chinese Lion, Globe). The first column is the 3D printed freeform interface. All touchpoints are depicted with black and all non-conductive parts of the objects is depicted with white. The second column is the circuit routing illustrating how the serpentine trace pattern is routed for each object. The third column is each object's RC delay graph. For each RC delay graph, the x-axis corresponds to the time elased since using the freeform interface. The y-axis depicts to the number of program loops that elapsed until reaching the microcontroller’s voltage threshold.}
    \label{fig:fab-objects}
\end{figure*}

\section{Applications}
\label{sec4:applications}
We demonstrate the applicability of our computational design pipeline with six freeform interfaces with different geometries. Besides the Stanford Bunny already mentioned, our other freeform interfaces include a MIDI Drumpad, a Hilbert Curve, a Chinese character (power), a museum artifact (Chinese lion), and a globe. 
\autoref{fig:fab-objects} shows the 3D printed model, its internal circuit, and its respective RC delay graph with the \singlewire\ connection. These objects were chosen to demonstrate our pipeline's ability to handle various geometries. The MIDI Drumpad is an example of a basic geometry; Hilbert Curve illustrates geometric curves; Power demonstrates working with a limited bounding volume; the Stanford Bunny and Chinese Lion are examples of freeform interfaces.
Except for the Chinese Character (power) and MIDI Drumpad, the remaining four models were selected from the Thingi10K dataset repository~\cite{zhou2016thingi10k}. See the supplemental materials to see the STL files for these freeform interfaces.
The freeform interfaces represent different numbers of touchpoints with the required \pipe\ length discussed in \autoref{fig:fab_scale}b.

\paragraph{MIDI Drumpad and Hilbert Curve.}
We envision our technique can be used to quickly prototype tangible interfaces that require many touchpoints. As discussed in \autoref{sec:related}, one of the limitations of directly embedding electronics is that it requires iterations of post-processing. To illustrate this vision, we provide two examples. \autoref{fig:fab-objects}c is a Hilbert curve with touchpoints. The Hilbert curve aims to show how our technique can enable touchpoints even for complex geometry. In contrast, \autoref{fig:fab-objects}b is a MIDI trackpad with a box shape.
The trackpad was modeled in a commercial CAD tool. The trackpad emphasizes how we can enable various touchpoints (i.e., 16 in this case). Both examples show how users can quickly create different tangible prototypes while minimizing the use of electronics and post-processing. 

\paragraph{Chinese Character (Power).}
Research highlights how tangible artifacts can increase learning engagement while also presenting the learning materials in a different manner~\cite{schneider2010benefits}. For example, \autoref{fig:fab-objects}d shows how the five touchpoints can be used to learn the sequential stroke order for the Chinese character for `power'. The touchpoints are embossed so a user can trace their finger along the surface of the character to learn how to write the character.

\paragraph{Cultural Heritage Artifacts.}
In most museum settings, visitors cannot directly touch historical artifacts. In these cases, visitors can only inspect the historical artifacts from afar. 3D printing technology can create replicas of cultural artifacts that visitors can engage with and provide a more interative way to learn~\cite{neumuller20143d}. While visitors may not be able to directly inspect historical artifacts, the replicas can have different embedded touchpoints that users can select for further inspection.  \autoref{fig:fab-objects}e shows where a visitor selects the Chinese lion's paw for closer details.

\paragraph{Globe.}
\autoref{fig:fab-objects}f shows a globe with six touchpoints to help learners identify the different continents.

\section{Evaluation}
\label{sec4:evaluation}
The goal of our evaluation is to enable a deeper understanding of this RC-delay capacitive sensing technique. 
Thus, our evaluations focus on the sensing technique rather than the pipeline as a tool. The theoretical, experimental, and computational investigation provides the groundwork for this objective, and are highly recognized methods for technical HCI work~\cite{hudson2014concepts}.

As a step toward this goal, we evaluate the efficacy of our approach with five technical evaluations and six applications.
We evaluate the practical constraints of our pipeline, specifically the number of touchpoints that we can fabricate while ensuring each touchpoint is distinguishable.
We perform a computational performance evaluation of the algorithms used in the automatic circuit design stage. 
We showcase six freeform interfaces made with our pipeline. 
We measured the signal-to-noise ratio of this sensing technique.
We conduct a user study to assess real-time recognition accuracy.
We conducted a computational experiment and mathematical analysis to evaluate the robustness of the \singlewire\ connection.

\subsection{Fabrication Scalability}
\label{sec4:scalability}
To determine the scalability of the capacitive touchpoints we can fabricate, we evaluate what the \textit{minimum} length of each \pipe\ (\SI{}{\milli\meter}) between two touchpoints should be to distinguish each selected point. 
Determining the minimum length of each \pipe\ between two touchpoints can infer the smallest possible volume of a freeform interface.
We do not evaluate the maximum size of a freeform interface as the maximum size is restricted by the build volume of a given 3D printer.

\paragraph{Double-Wire Connection.} We first consider the \doublewire\ connection condition. 
Based on \autoref{sec:one-vs-two}, the time difference required to reach a microcontroller’s logic threshold voltage when touching $p$th and $(p+1)$th touchpoint can be written as: ${t_\mathrm{thres}}_{p+1} - {t_\mathrm{thres}}_{p} = c r_{p+1} \ln ( v_\mathrm{in} / (v_\mathrm{in} - v_\mathrm{thres}) )$.
To analyze ${t_\mathrm{thres}}_{p+1} - {t_\mathrm{thres}}_{p}$, we place three assumptions that would represent common use:
\begin{itemize}
    \item $c=100$pF, as a representative capacitance for a human body when selecting a touchpoint~\cite{esd2010fundamentals}.
    \item $v_\mathrm{in}=5$V and $v_\mathrm{thres}=2.5$V, following the technical specifications of the Arduino UNO R4. 
    \item \si{{5}\micro\second} as the minimum value required for ${t_\mathrm{thres}}_{p+1} - {t_\mathrm{thres}}_{p}$, which corresponds to 240 clock cycles of the Arduino UNO R4's CPU.
\end{itemize}
From these conditions, we can derive $r_{p+1} \geq $\si{35\kilo\Omega}.

We need to determine the minimum length of a \pipe\ (3D pipe) that can house conductive traces with over \si{35\kilo\Omega}.
To determine such length, we introduce the following technical assumptions:
\begin{itemize}
    \item The use of Snapmaker J1S, a 3D printer with a 0.4 \si{\milli\meter} nozzle (standard for consumer FDM 3D printers)
    \item Using a \si{0.4\milli\meter} nozzle, the thickness of the conductive trace for the $xy$-plane is set to \si{{0.8}\milli\meter}; the ray and layer margins are set to \si{{1.2}\milli\meter} (refer to \autoref{sec:circuit-embedding}).
    \item Protopasta’s conductive PLA (\si{1.75\milli\meter}) as the conductive filament~\cite{protopasta}.
\end{itemize}

We measured the resistance of a conductive trace per length along the $xy$-plane and $z$-direction using Protopasta's conductive PLA. The results are \si{256\Omega/\milli\meter} for the $xy$-plane and \si{1013\Omega/\milli\meter} for the $z$-direction (\autoref{appendix-c:resistance-test}).
We focus only on the conductive traces on the $xy$-plane because our circuit embedding mainly relies on traces on the $xy$-plane  (\autoref{sec:circuit-embedding}).
Thus, to achieve \si{35\kilo\Omega}, we need to print approximately \si{137\milli\meter} of a conductive trace along the $xy$-plane.
This length can be achieved by drawing the conductive trace within a \pipe\ that has a \si{5\milli\meter} diameter (our pipeline's default value for the \pipes) and \si{9\milli\meter} length.
This result infers the minimum length of a conduit should be  \si{9\milli\meter} between each pair of touchpoints. 

Note that different technical assumptions can lead to different results.  
One significant but easily changeable assumption is the minimum value required for ${t_\mathrm{thres}}_{p+1} - {t_\mathrm{thres}}_{p}$ (i.e., \si{{5}\micro\second}). 
We consider \si{{5}\micro\second} to be a relatively safe value corresponding to over 200 clock cycles for the Arduino UNO 4. The value provides enough of a buffer to account for errors in the conductive trace's resistance (e.g., due to poor 3D printing precision).
However, if a user can confirm one's 3D printing errors are small (e.g., high-precision printing), the minimum required time delay difference can be radically reduced (e.g., \si{{1}\micro\second}).
This condition significantly reduces the required horizontal length of a conductive trace for each \pipe\ (e.g., from  \si{137\milli\meter} to  \si{27\milli\meter}). 

\begin{figure}[tb]
	\centering
	\captionsetup{farskip=0pt}
    \subfloat[Required resistance]{
        \includegraphics[height=0.25\linewidth]{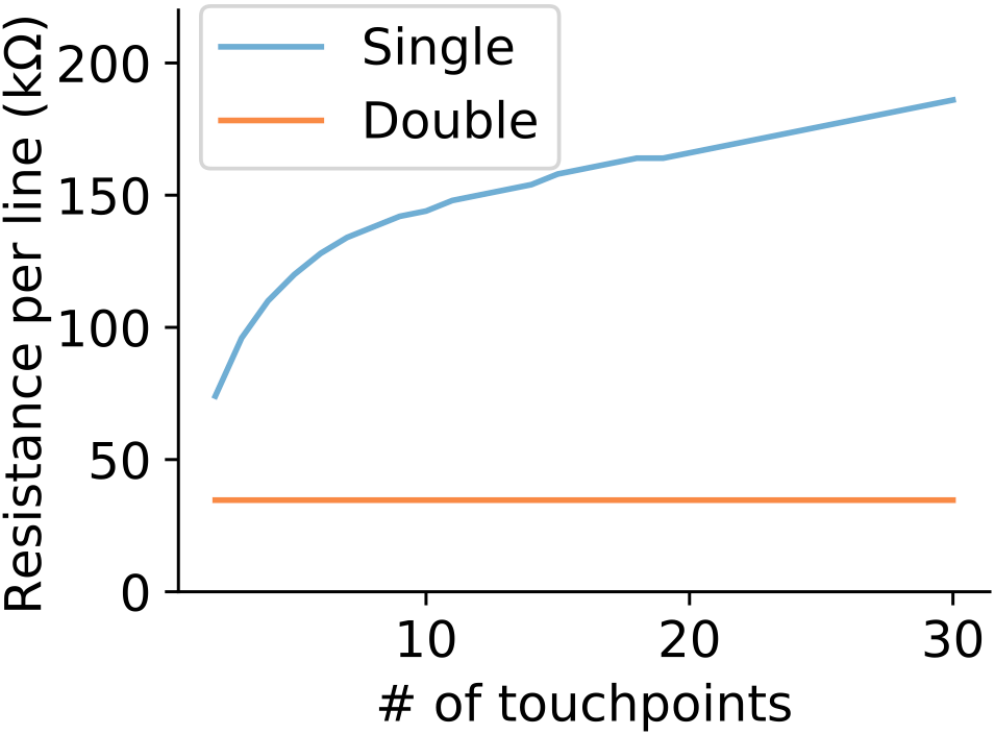}
        \label{fig:fab_scale_resistance}
    }
    \hspace{5pt}
    \subfloat[Required \pipe\ length]{
        \includegraphics[height=0.25\linewidth]{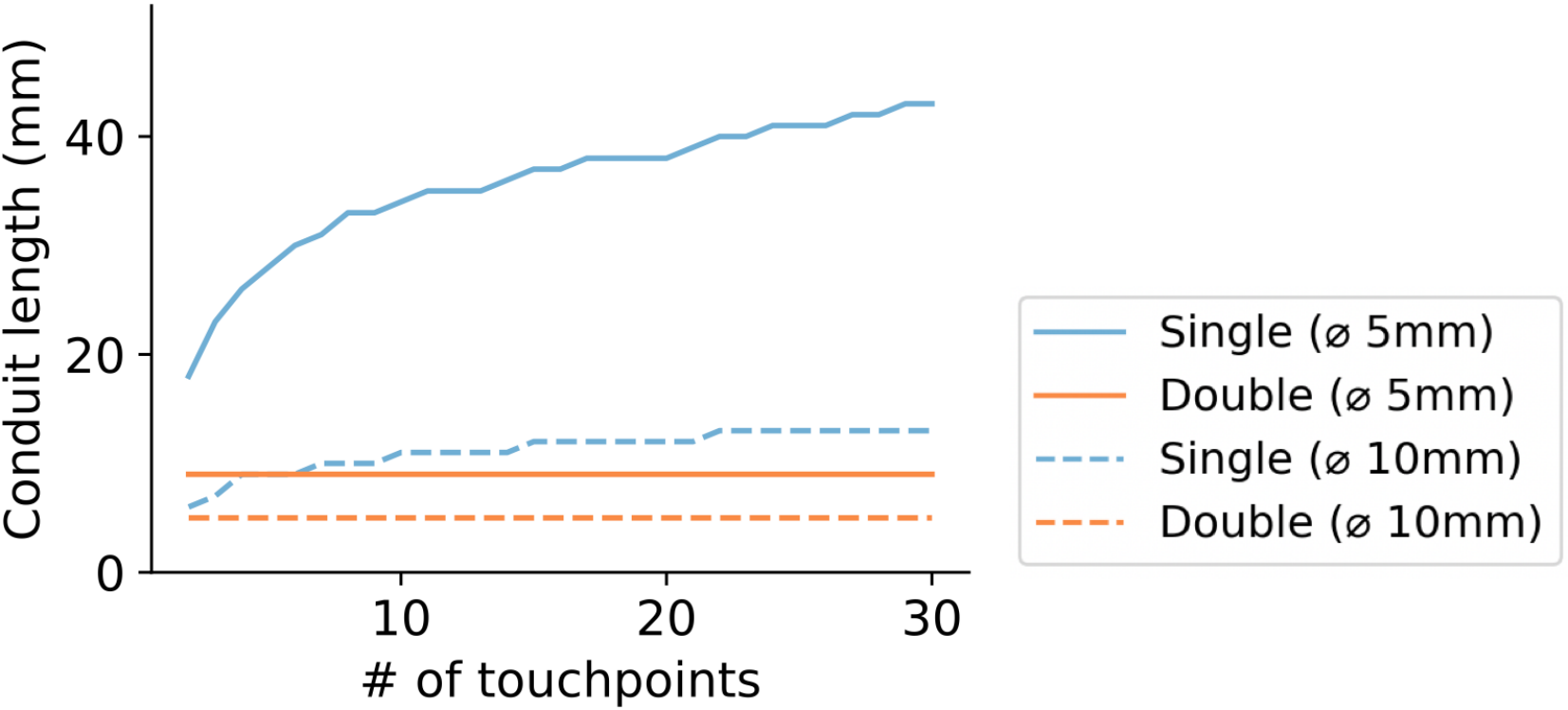}
        \label{fig:fab_scale_length}
    }
    \caption{Fabrication scalability evaluation of the single- and double-wire connections with different numbers of touchpoints. In (b), we assume that a \pipe\ has a fixed diameter of \si{5\milli\meter} (our pipeline's default) or \si{10\milli\meter}.}
    \Description{
     Two line graphs indicating the fabrication scalability of the single-wire and double-wire connections with different number of touch points. The x-axis ranges from 0 to 30 and is labeled "Number of Touchpoints" for both line graphs. The y-axis ranges from 0 to 200 and is labeled "Resistance per line (kilo ohm)" for the first line graph and "Conduit length" for the second line graph. Across both line graphs, the single-wire condition illustrates a positive trend as the number of touchpoint increases. Similarly, across both line graphs, the double-wire connection condition illustrates a flat line at a value slightly less than 50. 
    }
	\label{fig:fab_scale}
\end{figure}

\paragraph{Single-Wire Connection.} 
To understand the fabrication scalability for the \singlewire\ connection, we place the same assumptions we listed for the \doublewire\ connection condition.
Similar to \autoref{sec:one-wire-optimization}, we apply the constraint of $r_2=\cdots=r_n=r$ and perform a grid search of $r$ and $r_1$ to find the minimum value of $r$ that satisfies ${t_\mathrm{thres}}_{p+1} - {t_\mathrm{thres}}_{p} \geq 5$\si{\micro\second} for all $p$.
Unlike the \doublewire\ connection, the required $r$ for the \singlewire\ connection varies based on the number of touchpoints. 
\autoref{fig:fab_scale_resistance} summarizes the value of $r$ depending on the different number of touchpoints. 
Following the same procedure as the \doublewire\ connection, we derive the minimum length of a \pipe\ with a \si{5\milli\meter} diameter, as shown in \autoref{fig:fab_scale_length}.
The derived minimum length for 20 touchpoints is \si{40\milli\meter}.
However, as shown with the dashed lines in \autoref{fig:fab_scale_length}, if we increase the diameter of the \pipe\ to \si{10\milli\meter} diameter, this minimum length of a \pipe\ can be reduced to \si{12\milli\meter}.
For the \singlewire\ connection, we can infer that (1) the required length of a \pipe\ between two touchpoints is longer than the \doublewire\ connection and (2) the required length follows a close-to-logarithmic function as we increase the number of touchpoints.
Therefore, the \singlewire\ connection places a stronger constraint to fabricate a small freeform interface with numerous touchpoints.

\subsection{Computational Performance}
\label{sec:perf_eval}
We conducted a performance evaluation of the algorithms used in the automatic circuit design stage (\autoref{sec3:auto-circuit-design}). 
We first analyzed the time complexity of the algorithms to uncover potential performance bottlenecks when dealing with complex or large 3D objects.
We then ran an experimental evaluation on different 3D models shown in \autoref{tab:comp_perf}. The results revealed the automatic circuit design stages can be completed between \si{16\second} and \si{240\second}.

\newcommand{\nTriangles}{T}
\newcommand{\nGridPoints}{G}
\newcommand{\nVoxels}{V}
\newcommand{\Volume}{U}
\newcommand{\nPipeTriangles}{S}
\paragraph{Time Complexity.}
The computationally demanding steps are the (1) voxelization of the freeform model (\autoref{fig:pathfinding}b), (2) Dijkstra's pathfinding (\autoref{fig:pathfinding}f), and (3) circuit embedding (\autoref{fig:components}b).

\sloppy{Our voxelization uses the implementation provided by PyVista~\cite{sullivan2019pyvista}, which checks whether each position of voxel grids is within an object surface.}
Thus, the voxelization has $\mathcal{O}(\nTriangles \nGridPoints)$ where $\nTriangles$ is the number of triangles constructing a surface and $\nGridPoints$ is the number of grid points.
Note that $\nGridPoints$ is roughly proportional to the number of the resulting voxels, $\nVoxels$ (i.e., $\mathcal{O}(\nTriangles \nGridPoints) \approx \mathcal{O}(\nTriangles \nVoxels)$).
For each pair of touchpoints, Dijkstra's path-finding algorithm is performed with the weighted graph of the trimmed voxel representation (\autoref{fig:pathfinding}-d).
In total, this pathfinding has $\mathcal{O}(\nTouchpoints \nVoxels \log \nVoxels )$ where $\nTouchpoints$ is the number of touchpoints. 
The circuit embedding finds a line segment for each ray on a \pipe\ using the space-filling algorithm.
With $\nPipeTriangles$ triangles on a \pipe's surface, the line segmentation can be performed with $\mathcal{O}(\nPipeTriangles \log \nPipeTriangles)$.
When several rays are generated to slice a \pipe\ (modeled as a 3D pipe) with a small margin, the circuit embedding for each \pipe\ has a time complexity of $\mathcal{O}(\Volume \nPipeTriangles \log \nPipeTriangles )$ where $\Volume$ is the volume of a 3D object. 
We apply the serpentine trace pattern only for low conductivity \pipes.  In total, the circuit embedding takes $\mathcal{O}(\nTouchpoints \Volume \nPipeTriangles \log \nPipeTriangles )$.
However, if we model a \pipe's surface with a fixed small number of triangles, we can simplify the complexity to $\mathcal{O}(\nTouchpoints \Volume)$.

In sum, the automatic circuit design involves $\mathcal{O}(\nTriangles \nVoxels)$ (i.e., voxelization), $\mathcal{O}(\nTouchpoints \nVoxels \log \nVoxels )$ (i.e., Dijkstra's), and  $\mathcal{O}(\nTouchpoints \Volume)$ (i.e., circuit embedding) computations.
The results highlight that the critical parameters for computation are $\Volume$ (volume), $\nTriangles$ (the number of triangles), $\nVoxels$ (the number of voxels), and $\nTouchpoints$ (the number of touchpoints).

\begin{table}[b!]
    \setlength{\tabcolsep}{2pt}
    \scriptsize
    \centering
    % \renewcommand{\arraystretch}{0.95} %remove later for CR
  % \small
  \centering
  \begin{tabular}{
                  >{\raggedright}
                  p{0.43in}
                  c
                  cccc
                  ccccc}
\toprule
    & \multicolumn{4}{c}{\textbf{Object Information}} &  
    & \multicolumn{5}{c}{\textbf{Completion Time (\si{\second})}}\\
    \arrayrulecolor{black!30}\cmidrule(l){2-5} \cmidrule(l){7-11}
    & vol (\si{\milli\meter^3}) & triangles & voxels & \# points &  & voxelize & Dijkstra & circuit embed & misc & total \\
    \arrayrulecolor{black!30}\midrule
    
    %% ------------------------------------------------------
    %% Bunny
    %% ------------------------------------------------------
    \raisebox{-\totalheight}{\includegraphics[width=0.07\textwidth]{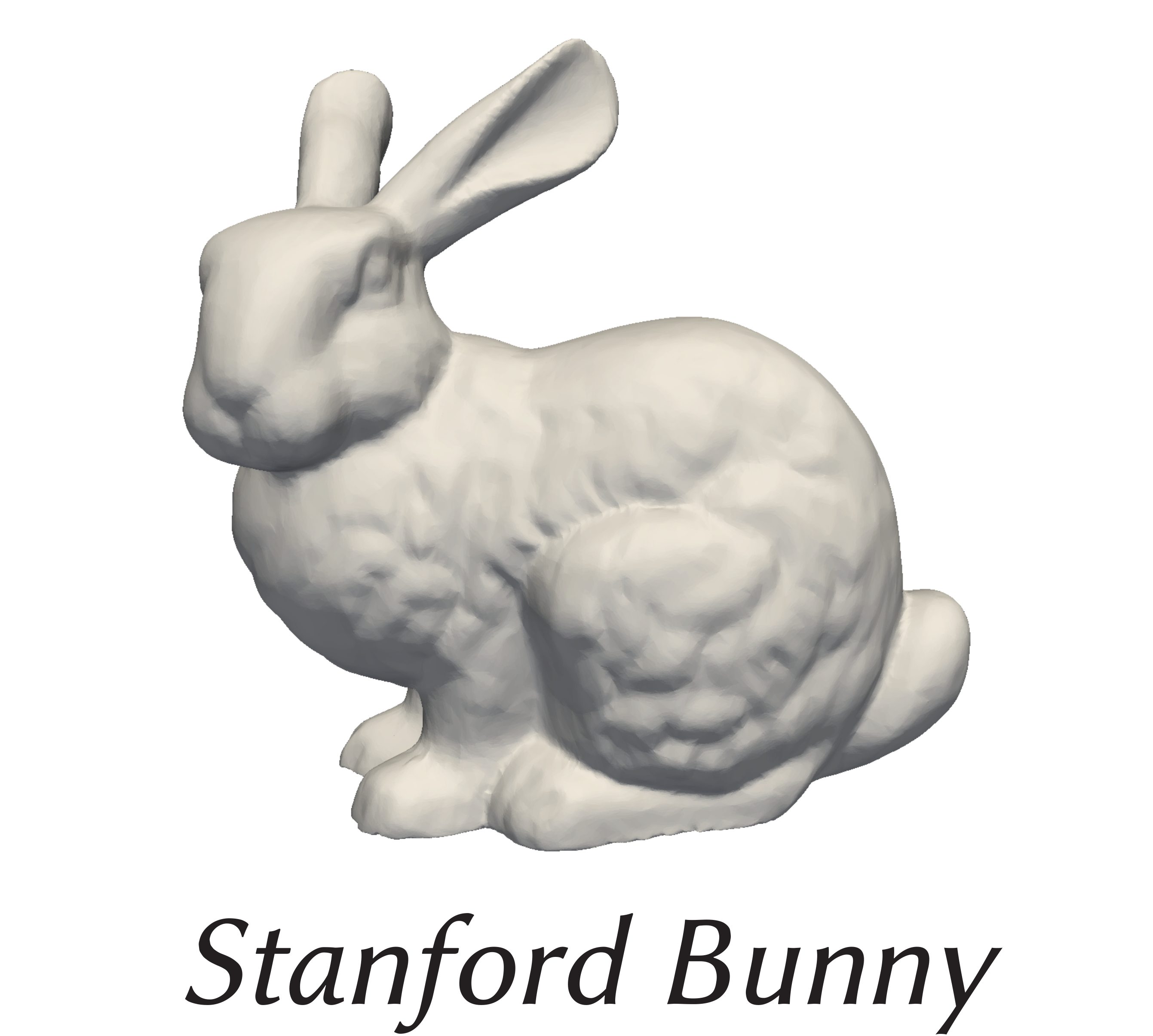}} & \multirow{8}{*}{302391} & \multirow{8}{*}{259898} & \multirow{8}{*}{444696} & \multirow{8}{*}{5} & &  \multirow{8}{*}{60} & \multirow{8}{*}{53} & \multirow{8}{*}{12} & \multirow{8}{*}{28} & \multirow{8}{*}{155}\\

    %% ------------------------------------------------------
    %% Pad4
    %% ------------------------------------------------------
   \raisebox{-\totalheight}{\includegraphics[width=0.07\textwidth]{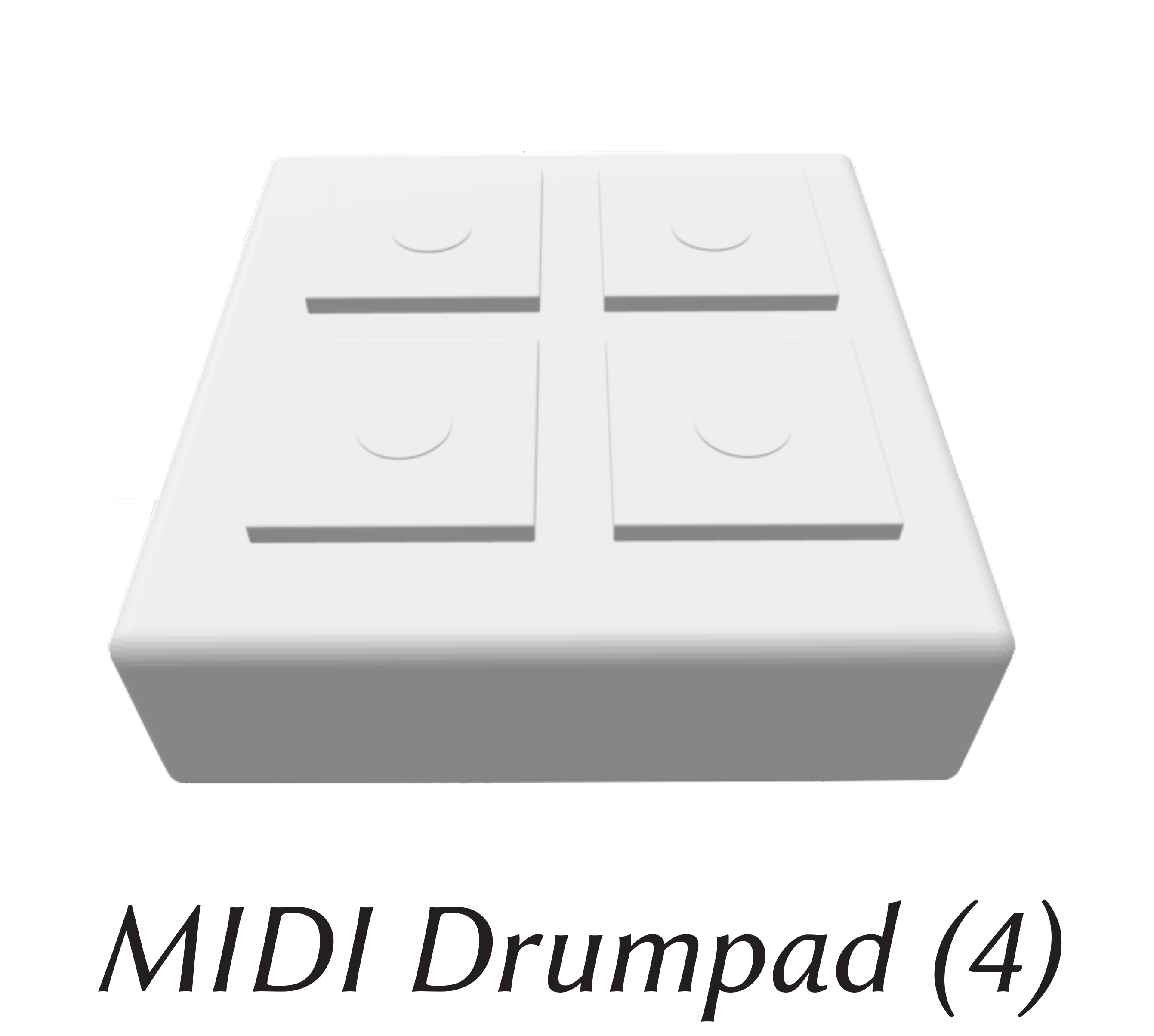}} & \multirow{7}{*}{78284} & \multirow{7}{*}{2116} & \multirow{7}{*}{50176} & \multirow{7}{*}{4} & &  \multirow{6}{*}{1} & \multirow{7}{*}{3} & \multirow{7}{*}{6} & \multirow{7}{*}{5} & \multirow{7}{*}{16} \\

    %% ------------------------------------------------------
    %% pPad9
    %% ------------------------------------------------------
\raisebox{-\totalheight}{\includegraphics[width=0.07\textwidth]{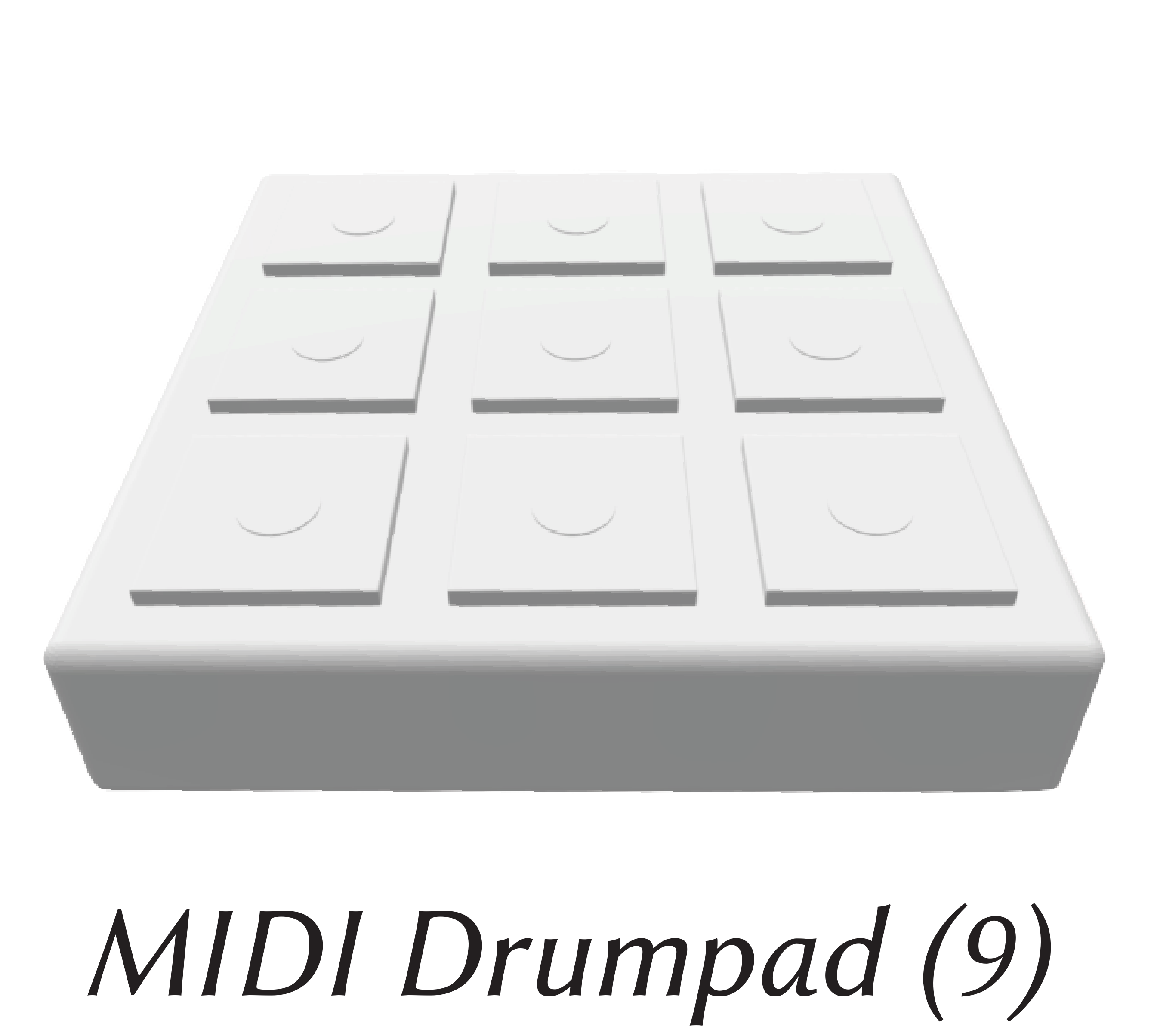}}& \multirow{8}{*}{241964} & \multirow{8}{*}{7336} & \multirow{8}{*}{155232} &\multirow{8}{*}{9} & &  \multirow{8}{*}{2} & \multirow{8}{*}{7} & \multirow{8}{*}{15} & \multirow{8}{*}{15} & \multirow{8}{*}{42} \\

    %% ------------------------------------------------------
    %% Pad16
    %% ------------------------------------------------------
\raisebox{-\totalheight}{\includegraphics[width=0.07\textwidth]{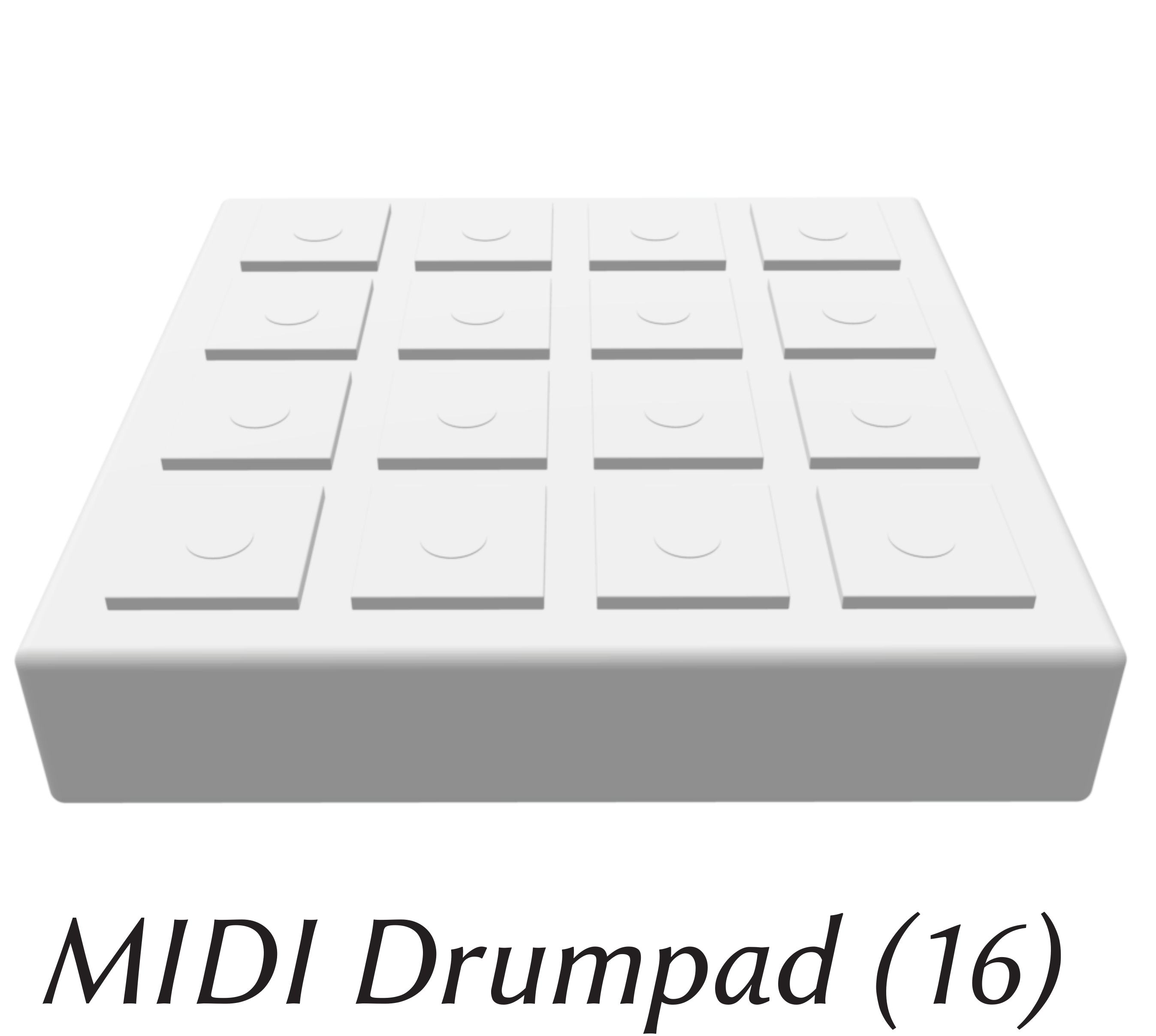}}& \multirow{8}{*}{458519} & \multirow{8}{*}{10316} & \multirow{8}{*}{320211} & \multirow{8}{*}{16} &  &  \multirow{8}{*}{5} & \multirow{8}{*}{20} & \multirow{8}{*}{149} & \multirow{8}{*}{44} & \multirow{8}{*}{223} \\

    %% ------------------------------------------------------
    %% Hilbert
    %% ------------------------------------------------------
\raisebox{-\totalheight}{\includegraphics[width=0.07\textwidth]{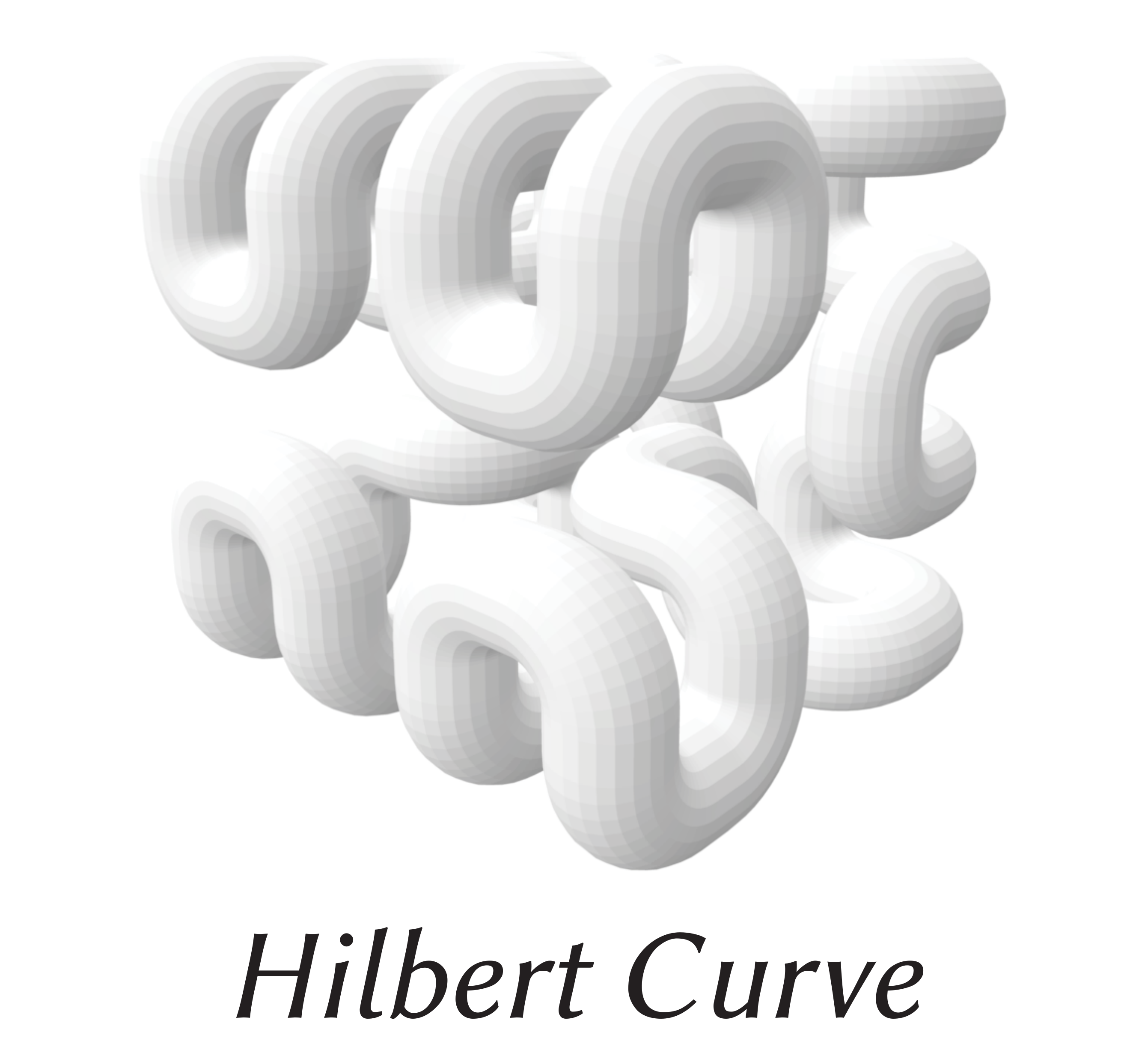}}& \multirow{8}{*}{173667} & \multirow{8}{*}{39936} & \multirow{8}{*}{285897} & \multirow{8}{*}{10} & &  \multirow{8}{*}{12} & \multirow{8}{*}{1} & \multirow{8}{*}{51} & \multirow{8}{*}{49} & \multirow{8}{*}{117} \\

    %% ------------------------------------------------------
    %% Globe
    %% ------------------------------------------------------
\raisebox{-\totalheight}{\includegraphics[width=0.07\textwidth]{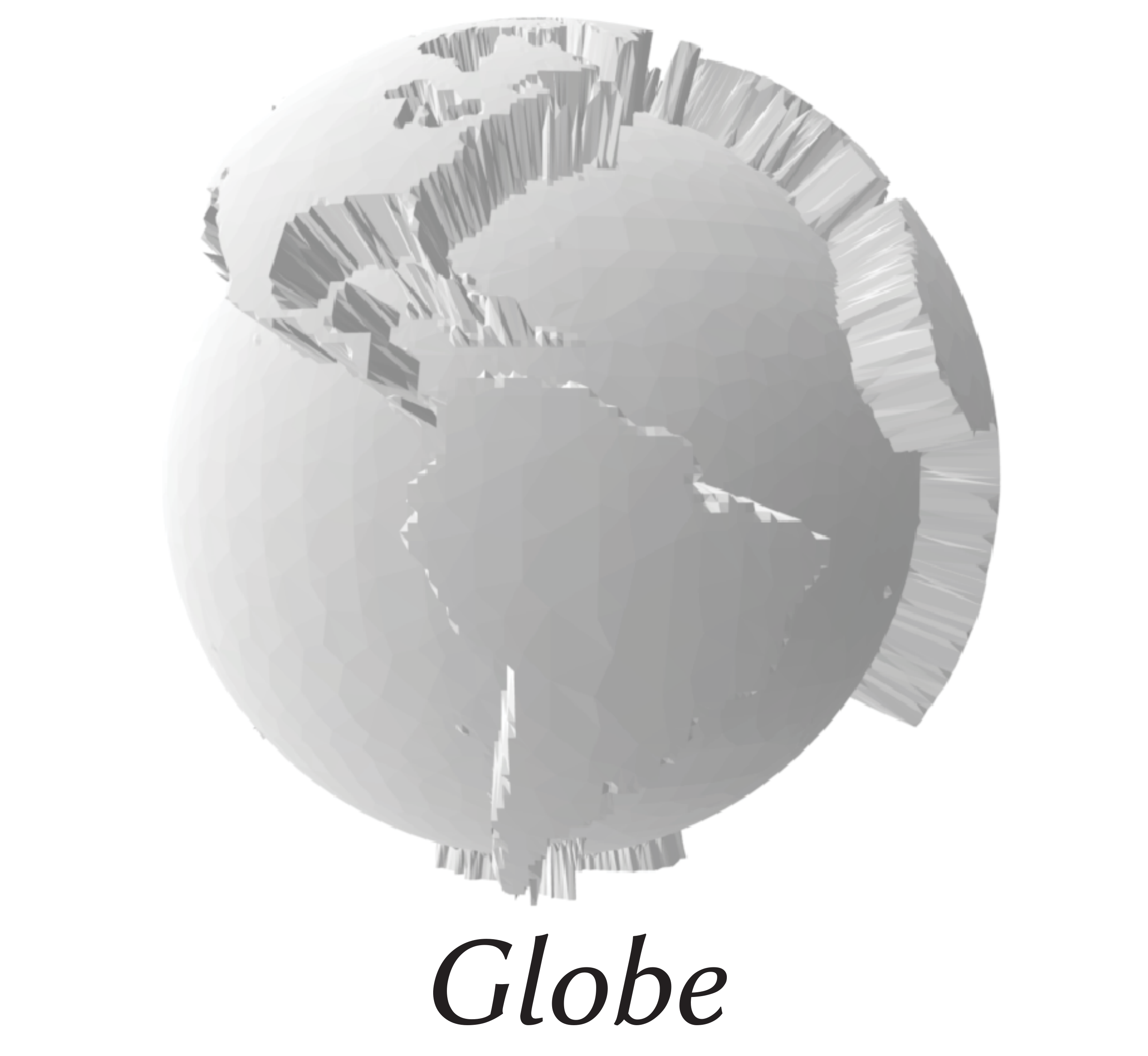}}& \multirow{8}{*}{317412} & \multirow{8}{*}{29472} & \multirow{8}{*}{325651} & \multirow{8}{*}{6} & &  \multirow{8}{*}{19} & \multirow{8}{*}{23} & \multirow{8}{*}{9} & \multirow{8}{*}{23} & \multirow{8}{*}{74} \\

    %% ------------------------------------------------------
    %% Lion
    %% ------------------------------------------------------
\raisebox{-\totalheight}{\includegraphics[width=0.07\textwidth]{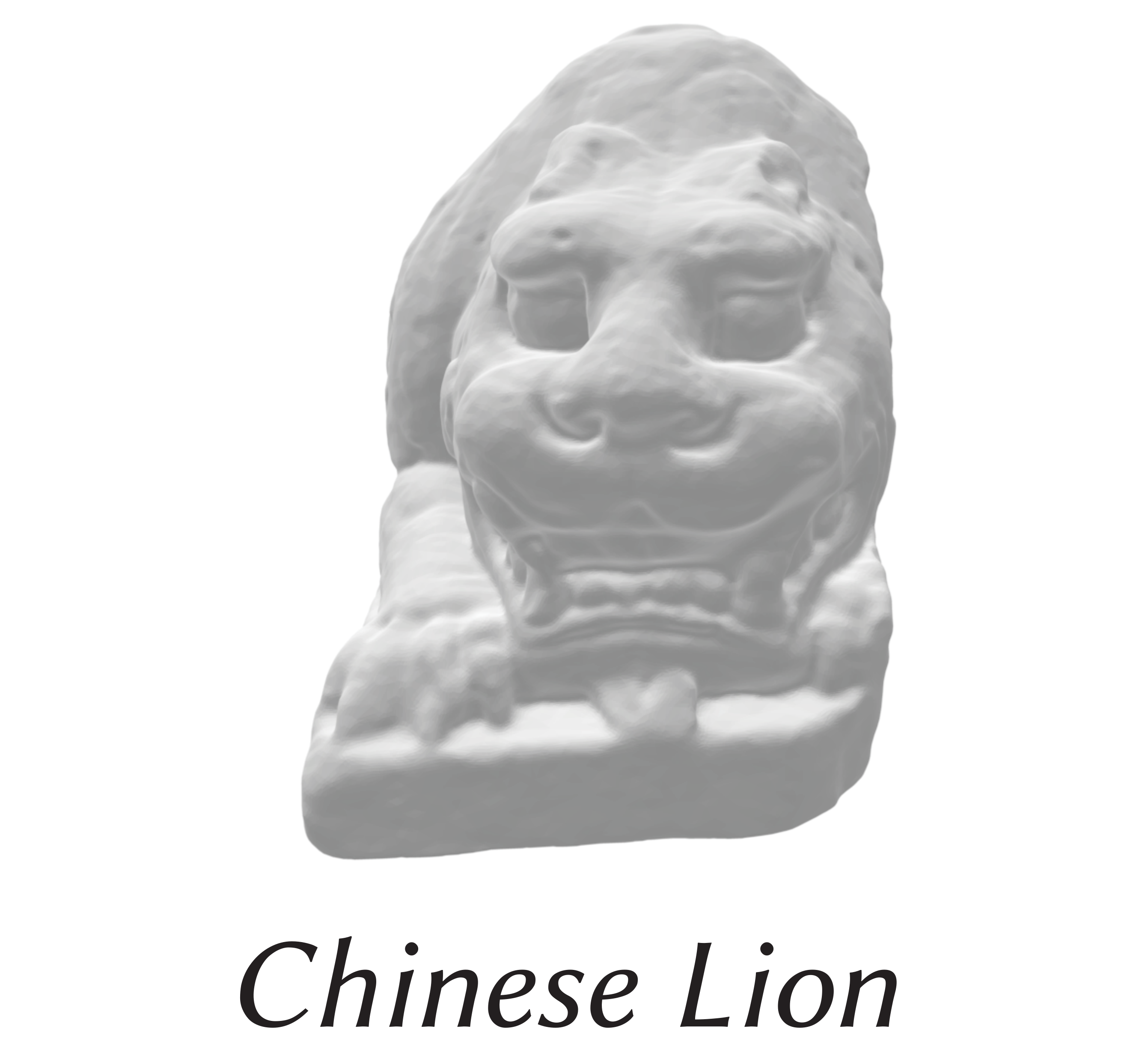}}& \multirow{8}{*}{613578} & \multirow{8}{*}{69994} & \multirow{8}{*}{384825} & \multirow{8}{*}{6} & &  \multirow{8}{*}{23} & \multirow{8}{*}{133} & \multirow{8}{*}{23} & \multirow{8}{*}{22} & \multirow{8}{*}{201} \\

    %% ------------------------------------------------------
    %% Power
    %% ------------------------------------------------------
    \raisebox{-\totalheight}{\includegraphics[width=0.07\textwidth]{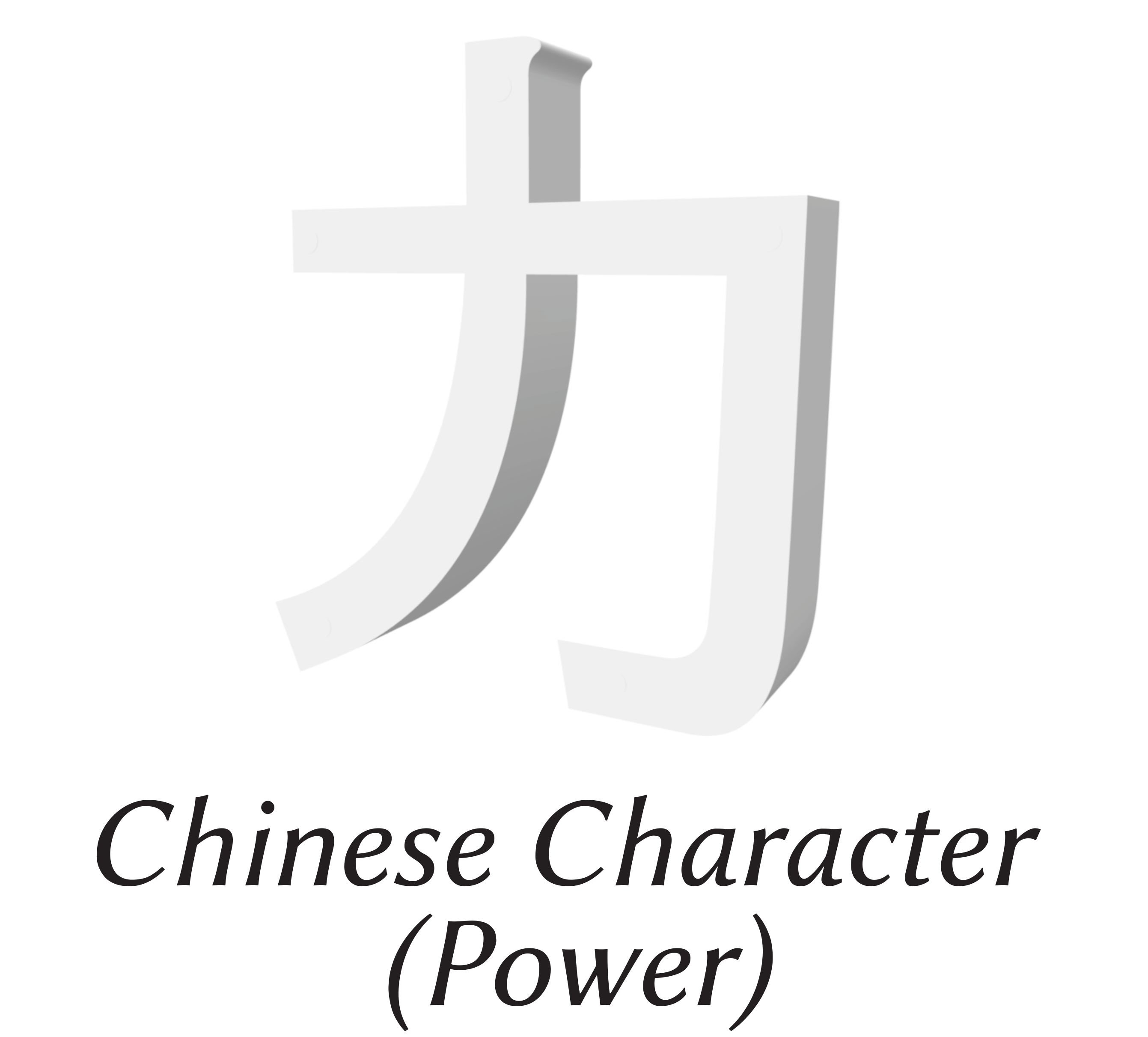}}& \multirow{8}{*}{132102} & \multirow{8}{*}{2256} & \multirow{8}{*}{295056} & \multirow{8}{*}{5} &  & \multirow{8}{*}{6} & \multirow{8}{*}{156} & \multirow{8}{*}{59} & \multirow{8}{*}{18} & \multirow{8}{*}{240} \\
\arrayrulecolor{black}\bottomrule

\end{tabular}
    \caption{Computational performance evaluation.}
    \label{tab:comp_perf}
\end{table} 

\paragraph{Experimental Evaluation.}
We used a MacBook Pro with 2.3 GHz 8-Core Intel Core i9 and 64 GB 2,667 MHz DDR4 (no GPU use). 
We collected and modeled eight 3D objects as seen in \autoref{tab:comp_perf}.
For each object, we ran the automatic circuit design stage five times and measured the average completion time. 
The breakdown and total completion times are shown in \autoref{tab:comp_perf}.
The miscellaneous steps shown in \autoref{tab:comp_perf} include the clipping touchpoints, the conversion from the trimmed voxel to the graph representation, and the resistance optimization. 
As expected from the time complexity analysis, completion time differs based on the number of triangles, voxels,  touchpoints, and model's volume.  
However, for the selected 3D models, the automatic circuit design is completed in less than 4 minutes for all objects.

\subsection{Signal-to-Noise Ratio}
Signal-to-Noise Ratio (SNR) is a quality metric that measures signal strength versus noise influence. This information can infer the likelihood of a false touch selection. 
SNR for capacitive sensing systems measures how robust the signals produced by the sensing technique (i.e., active signal) are compared to disturbances of background noise (i.e., inactive signal). 

\begin{table}[t]
  \centering
  \small
  \begin{tabular}{m{0.1cm}
                  m{3.2cm} 
                  m{1cm}
                  m{1cm}
                  m{1cm}
                  }
\toprule
    \multicolumn{2}{l}{\multirow{2}{*}{\textbf{Object \& Wiring Condition}}} 
    & 
    % \multirow{2}{*}{\textbf{Touchpoints}} &
    \multirow{2}{*}{\textbf{Trial 1}}
    & 
    \multirow{2}{*}{\textbf{Trial 2}} 
    & 
    \multirow{2}{*}{\textbf{Trial 3}}\\\\
     \arrayrulecolor{black!30}\midrule

    %% ------------------------------------------------------
    %% Stanford Bunny
    %% ------------------------------------------------------
    \multirow{2}{*}{\textbf{Stanford Bunny  ($n=5$)}} & ~
     & ~ & 
    ~ & 
    ~\\
     
     ~ &  \multirow{2}{*}{Double Wire} &
     \multirow{2}{*}{49.993} & 
     \multirow{2}{*}{55.179} & 
     \multirow{2}{*}{45.654} \\
    
    ~ &  \multirow{2}{*}{Single Wire} &
     \multirow{2}{*}{369.131} & 
     \multirow{2}{*}{323.638} & 
     \multirow{2}{*}{408.109} \\\addlinespace[1em]
     \midrule

    % %% ------------------------------------------------------
    % %% Hilbert Curve
    % %% ------------------------------------------------------
       \multirow{2}{*}{\textbf{Hilbert Curve ($n=10$)}} & ~
     & ~ & 
    ~ & 
    ~\\
     
     ~ &  \multirow{2}{*}{Double Wire} &
     \multirow{2}{*}{18.779} & 
     \multirow{2}{*}{17.930} & 
     \multirow{2}{*}{17.921} \\
    
    ~ &  \multirow{2}{*}{Single Wire} &
     \multirow{2}{*}{131.987} & 
     \multirow{2}{*}{132.576} & 
     \multirow{2}{*}{124.476} \\\addlinespace[1em]
     \midrule

    % %% ------------------------------------------------------
    % %% Drumpad
    % %% ------------------------------------------------------
       \multirow{2}{*}{\textbf{Drumpad 16  ($n=16$)}} & ~
     & ~ & 
    ~ & 
    ~\\
     
     ~ &  \multirow{2}{*}{Double Wire} &
     \multirow{2}{*}{49.930} & 
     \multirow{2}{*}{44.784} & 
     \multirow{2}{*}{49.263} \\
    
    ~ &  \multirow{2}{*}{Single Wire} &
     \multirow{2}{*}{22.432} & 
     \multirow{2}{*}{22.099} & 
     \multirow{2}{*}{25.876} \\\addlinespace[1em]

\arrayrulecolor{black}\bottomrule

\end{tabular}
  \vspace{-1em}
  \caption{Signal-to-noise ratio for Stanford Bunny, Hilbert Curve, and MIDI Drumpad (16) under the two wiring conditions: single wire and double wire. $n$ represents the number of touchpoints. }
  \label{tab:snr}
\end{table}

We employed a similar approach to past work~\cite{palma2024capacitive}, where we also computed SNR by repeatedly touching objects with the index finger. Three objects (Stanford Bunny, Hilbert Curve, and MIDI Drumpad 16) were chosen based on the different number of touchpoints and geometry complexity. We touched all touchpoints for a given object (refer to \autoref{fig:fab-objects} to see touchpoint placements). For each touchpoint, we adhered to a three-part process that lasted 9 seconds. First, we did not touch for 3 seconds. Then we touched the designed touchpoint for 3 seconds and lastly let go for 3 seconds. This process was repeated for 3 trials. During this process, we measured the raw capacitive values from the Arduino Uno R4. 
From these raw values, we used Davidson's proposed formula~\cite{davison2010techniques} to compute SNR  (\autoref{eq:snr}). $\mu_U$ is the mean value when the touchpoint is not pressed. $\mu_P$ is the mean value the touchpoint is pressed. $\sigma_U$ is the standard deviation of values when the touchpoint is not pressed.

\begin{equation}
    \label{eq:snr}
    SNR =\frac{\left|\mu_U-\mu_P\right|}{\sigma_U}
\end{equation}

Traditionally,  $\mu_U, \sigma_U$ for most capacitive sensing systems (e.g., \cite{palma2024capacitive,pourjafarian2019multi}) represents the inactive signal (i.e., background noise). These systems cast a binary judgment of whether a touchpoint has been selected or not. However, in our case, any other touchpoint besides the target touchpoint is also considered background noise. Our capacitive sensing technique relies on a \textit{categorical} judgment of determining which touchpoint is selected based on the RC Delay. As such, a system can wrongly judge a touchpoint selection if there is not enough difference in the RC delays among the touchpoints. Hence, for our SNR calculations, we computed a pairwise calculation ($n \times n$ matrix) between the target touchpoint ($\mu_P$) and all of the other touchpoints ($\mu_U, \sigma_U$). We report the \textit{minimum} SNR value from this pairwise computation to illustrate the smallest gap between a pair of touchpoints (\autoref{tab:snr}). See the supplemental material for the full pairwise computations. Davidson notes that the SNR threshold should at a minimum be 7, but ideally at least 15 for for robust sensing in real-world applications~\cite{davison2010techniques}.
Our results highlight how all combinations of the objects and wiring conditions satisfy this threshold, providing a high-level of reliability.

\subsection{Recognition Accuracy}
\label{sec:recog_acc}
We evaluate two of the freeform interfaces discussed in \autoref{sec4:applications}.
We conducted a controlled study to measure the real-time recognition accuracy of the freeform interfaces. The \singlewire\ and \doublewire\ connections are our independent variables. The accuracy of recognizing touchpoints is our dependent variable. Based on our discussion in \autoref{sec:one-wire} and insights from \autoref{sec4:scalability}, we expect the \singlewire\ connection to be more sensitive if a freeform interface has too many touchpoints (i.e., unable to distinguish between touchpoints). To validate this hypothesis, we measure our approach's real-time recognition accuracy with 8 different objects.

\subsubsection{Objects} Eight objects were selected based on the different number of touchpoints. The objects were generated from four different models: MIDI Drumpad with 4 keys \scshape{(PAD\_4)}, \normalfont MIDI Drumpad with 9 keys \scshape{(PAD\_9)}, \normalfont MIDI Drumpad with 16 keys \scshape{(PAD\_16)}, \normalfont and a Hilbert Curve with 10 touchpoints \scshape{(HILBERT\_10)}. \normalfont 
Each model was printed twice to represent the \singlewire\ and \doublewire\ connections. Each touchpoint was labeled with a number to indicate its touchpoint ID. 

We chose the MIDI Drumpads and Hilbert curve to represent both simple and complex geometry. 
The MIDI Drumpad's overall geometry as a box is simple enough that it would be easy for users to recognize and select the touchpoints. The simplicity in the geometry also lends well to how the same design can be easily scaled accordingly to generate different numbers of touchpoints.  However, the design of the MIDI Drumpad limits assessing whether geometric complexity can also affect real-time recognition accuracy. As a result, we include a Hilbert curve with 10 touchpoints. We anticipate the Hilbert Curve is one of the most complex geometries that fits our modeling criteria.

\subsubsection{Protocol} We recruited 10 participants for the study. Each study session was held in the same location to account for environmental factors. 
The 3D printed object is connected to an Arduino UNO 4 microcontroller, which is connected to a laptop. The laptop was powered by a wall outlet (earth ground). Participants were first given an overview of our capacitive sensing mechanism, and the experimenter demonstrated the sensing of touchpoints using the Stanford Bunny as an example. 

After the demonstration, participants began the formal study. We alternated the order of the two conditions (i.e., \singlewire\ and \doublewire\ connection) per participant. For each object, the participants were given verbal instructions for the calibration process (\autoref{sec3:physical-assembly}). They were instructed on the order of the touchpoints (left to right; row by row). For the MIDI pads, they held onto each touchpoint for \si{7\second} to account for fluctuation and noise. For the Hilbert Curve, participants were given \si{3\second} additional seconds to find the appropriate touchpoint because of the object's complex geometry. After calibration, the performance evaluation started. For each object, we randomly generated the order of the touchpoint ID the participants should touch (i.e., target touchpoint). 
Three trials were performed for each touchpoint per object.

\subsubsection{Data Collection and Analysis} For each object, we collected the participant's calibration and performance data with a signal processing library~\cite{bae2023computational}. The performance data is a time series with $\sim$64 samples per second. Each instance has a timestamp, the target touchpoint, and the classified touchpoint.
For each touchpoint, we trim the first 3.5 seconds during our analysis to account for the transition time between touchpoints or look-up time to find the target touchpoint.
This decision is based on our pilot study where we observed that a participant generally took about 2--3 seconds for transitions and look-up time. We added one additional second to provide a safe margin.
Within the remaining time, we set a 70\% threshold to determine the dominant touchpoint classification. The classification can fluctuate between two or more different touchpoint IDs if the RC delays are not distinct enough. If there is no dominant value within those remaining seconds, we consider the result to be \textit{not recognized} (i.e., no convergence to a value).

\begin{figure}[t!]
    \centering
    \includegraphics[width=\linewidth]{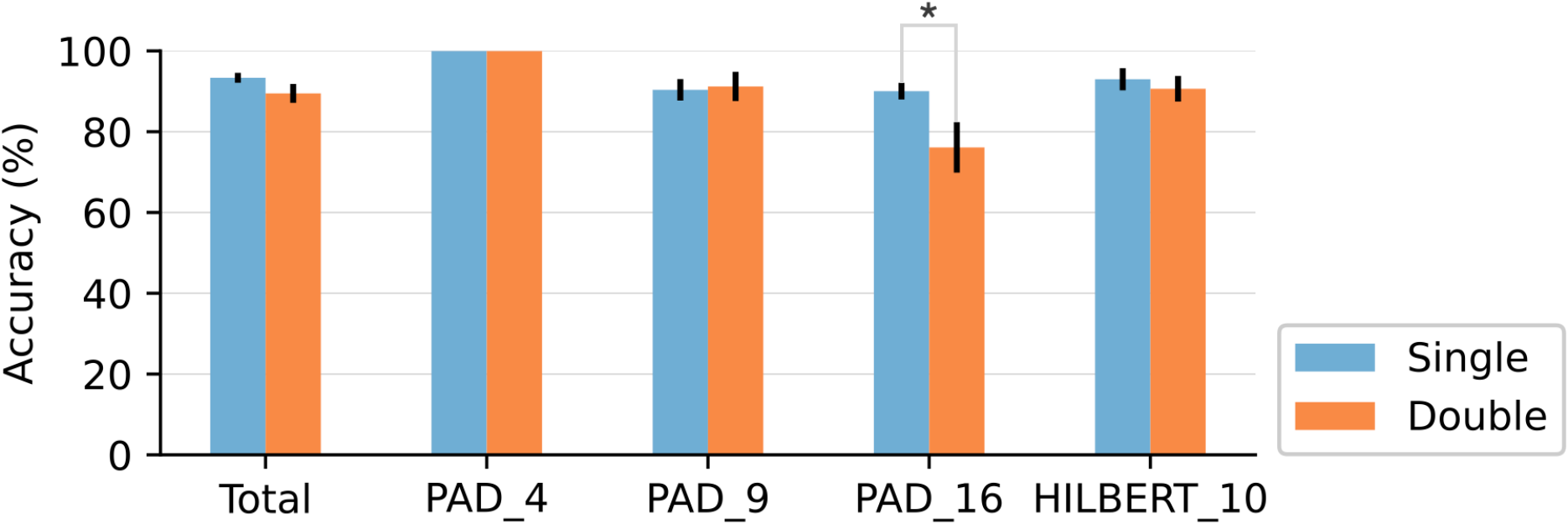}
    \caption{Results from the real-time recognition accuracy. Bar charts show the average accuracy across $n = 10$ participants with a 70\% threshold. Error bars represent standard errors. 
    There is a significant difference between the \singlewire\ and \doublewire\ connections for \scshape{PAD\_16}}.
    \Description{A bar chart depicting the results from the real-time recognition accuracy across 10 participants. The bar chart shows the results difference between the single-wire and double-wire connection. The average is 91\% for a 70\% threshold. Error bars represent standard errors. There is a significant difference between the single-wire and double-wire connections for PAD16.}
    \label{fig:overview-accuracy}
\end{figure}

\begin{figure}[t!]
    \centering
    \includegraphics[width=\linewidth]{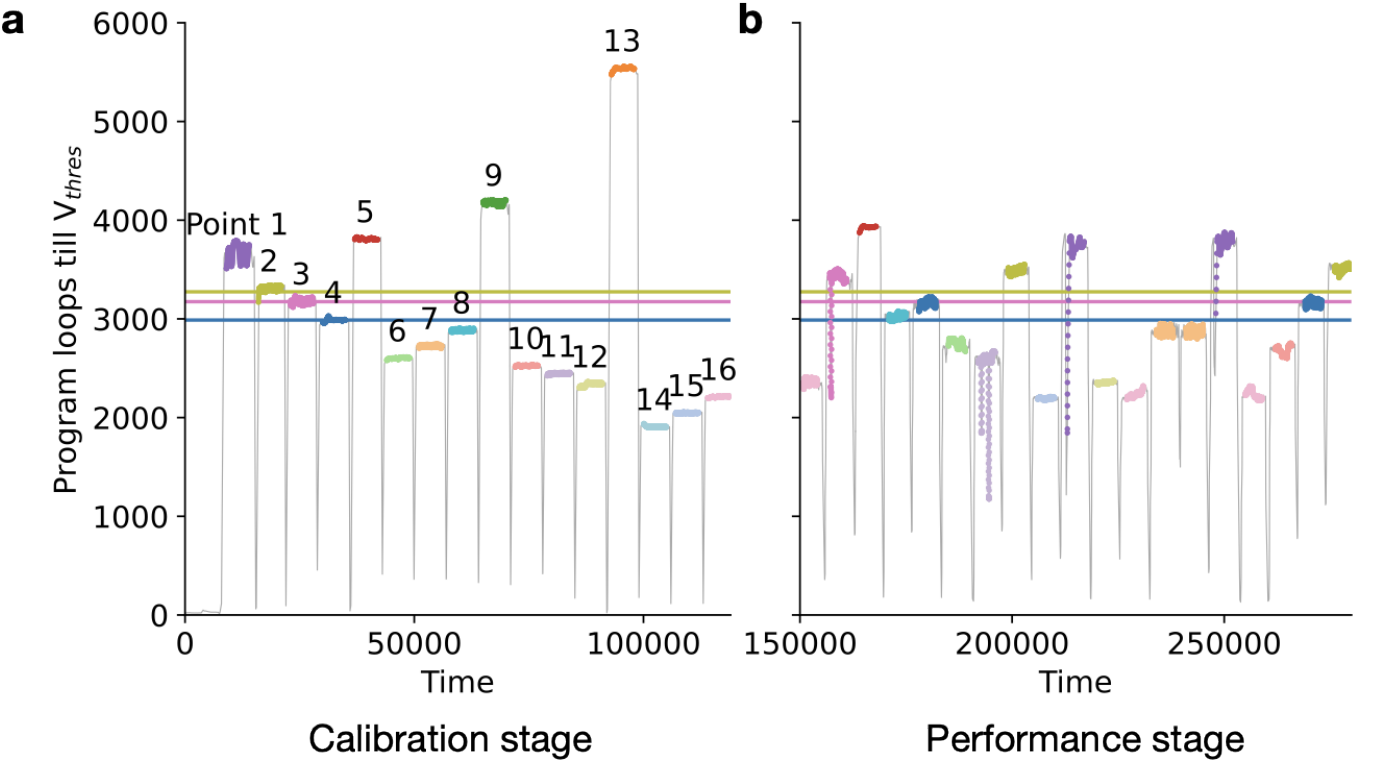}
    \caption{The RC delay graphs from P5's calibration and performance data for \textsc{PAD\_16} with the \doublewire\ connection. 
    Each graph is colored based on the target touchpoint. 
    The colored horizontal lines highlight the mean RC delays measured during the calibration stage (yellow: Point 2, pink: Point 3, blue: Point 4).
    In (b), we observe that the RC delay of each target node is shifted upwards.
    For example, a horizontal line shows how Point 4 (blue) in the performance data corresponds to Point 3 (pink), resulting in a misrecognition.}
    \Description{Two RC Delay Graphs from P5's calibration and performance data for PAD_16 for the double-wire condition. Each peak in the RC Delay is colored based on the target touchpoint. Three horizontal lines cross both RC delay graphs to illustrate how the mean RC Delays have shifted from the calibration stage.}
    \label{fig:pad16}
\end{figure}

\subsubsection{Findings}
Overall, our method yields an average accuracy of 91.42\% (SE=1.329) across the 8 freeform interfaces. 
\autoref{fig:overview-accuracy} shows that 6 out of 8 freeform interfaces achieve real-time accuracy of 90\% or higher. Most objects in the \singlewire\ connection demonstrated higher accuracy compared to their \doublewire\ connection.

Overall mean accuracies are 93.35\% (SE=1.223) for the \singlewire\ connection and 89.49\% (SE=2.339) for the  \doublewire\ connection.
However, a Wilcoxon signed-rank test did not show a statistically significant difference between the two conditions' accuracies ($Z = 1.376$, $p = 0.08445$). 

The model that had the greatest difference was \scshape{PAD\_16}. \normalfont A Wilcoxon signed-rank exact test showed there is a significant difference ($p = 0.03723$) in the accuracies between the \singlewire\ and \doublewire\ connections. 
Only one participant had an accuracy of less than 80\% with \scshape{PAD\_16} \normalfont in the \singlewire\ connection.  In contrast, five participants had an accuracy of less than 80\% with the \doublewire\ connection. 

A reason for the misrecognition errors is due to participants' RC delays shifting from the calibration stage to the performance stage.
For example, we examined the collected data from P6 who scored only 33\% accuracy for \textsc{PAD\_16} in the \doublewire\ connection.
\autoref{fig:pad16} shows how P6's performance data for all touchpoints has shifted from their calibration stage.
Consequently, most of P6's touchpoint selections were misclassified (e.g., Point 4 in \autoref{fig:pad16}a was misclassified as Point 3 during the performance stage).

\subsection{Robustness to Capacitance Shift}
\label{sec8:robustness}
The results in \autoref{sec:recog_acc} indicate that a shift in capacitance, $c$, can significantly influence the recognition accuracy.
Given how we observed an overall higher accuracy of the \singlewire{} connection than \doublewire{}, we hypothesize that the \singlewire{} connection is more robust to a capacitance shift.
To validate this hypothesis, we conduct a computational experiment perturbing the capacitance and perform a mathematical analysis.

\subsubsection{Computational Experiment}
\sloppy{Utilizing a circuit simulator, Lcapy~\cite{hayes2022lcapy}, we compare the robustness of the \doublewire's and \singlewire's connections to a capacitance shift.}
The circuit models correspond to the 8 freeform interfaces evaluated in \autoref{sec:recog_acc}.

We first resembled the calibration stage (\autoref{sec3:physical-assembly}) as follows.
With the circuit models, for each touchpoint, $p$, we derived the time required to reach a microcontroller's logic threshold voltage (refer to \autoref{eq:time-two-wires-approx} and \autoref{eq:time-one-wire-approx}) by setting $c=$\si{100\pico\farad} as a representative capacitance~\cite{esd2010fundamentals}.
We denote the derived time as $\TCalib{p}$.

To mimic the recognition stage, we perturbed $c$ from $\si{100\pico\farad}$ by randomly sampling $c$ from a Gaussian distribution with $\mu=100pF$ and $\sigma$.
We evaluated different $\sigma$ values ranging from $\si{0\pico\farad}$ to $\si{5\pico\farad}$.
For each $\sigma$, we sampled $c$ 100 times, which corresponds to testing the recognition with 100 participants.
For each sampled $c$, we derived the corresponding time required to reach the threshold voltage, $\TRecog{p}$, for all touchpoints.
Then, for each touchpoint, we judged whether the touch recognition was correct if $\arg\min_{q\in \{1, \cdots, \nTouchpoints \}} \lvert \TRecog{p} - \TCalib{q} \rvert = p$ (note: $\nTouchpoints$ is the number of touchpoints).
We computed the average accuracy for the 100 sampled $c$ values and $\nTouchpoints$ touchpoints for all 8 freeform interfaces.

\begin{figure}[t!]
    \centering
    \includegraphics[width=\linewidth]{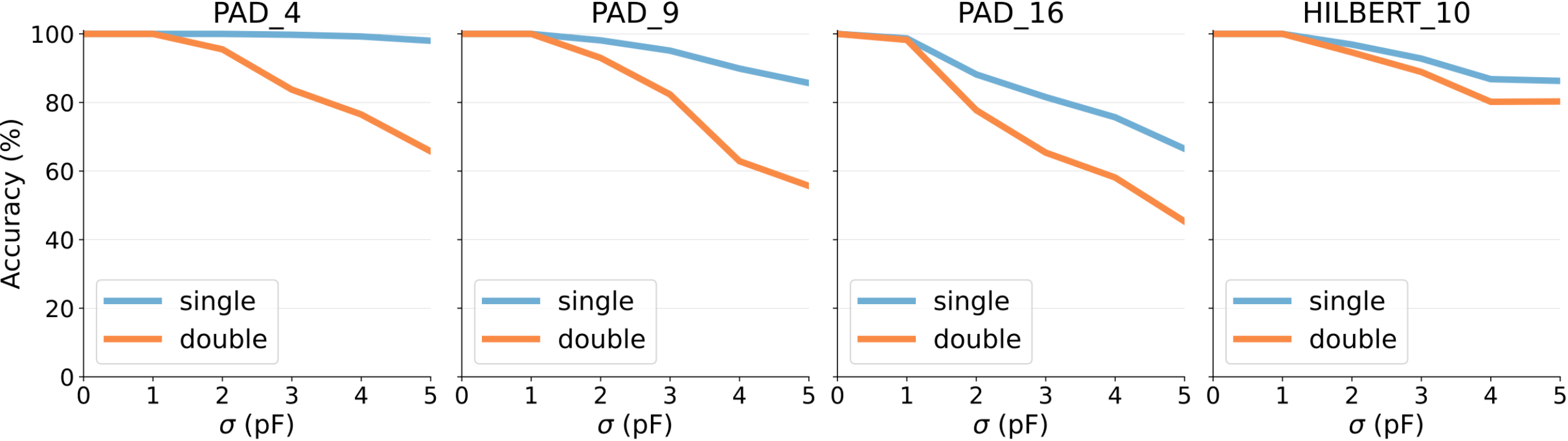}
    \caption{Results from the capacitance perturbation simulation for the freeform interfaces. $\sigma$ represents the strength of the perturbation (specifically, capacitance values are sampled from a Gaussian distribution with $\mu=100pF$ and $\sigma$). The \singlewire\ connection is more robust to the capacitance shift than the  \doublewire\ connection.}
    \label{fig:perturb}
\end{figure}

\autoref{fig:perturb} summarizes the results: the \singlewire\ connection is significantly more robust to a capacitance shift than the \doublewire\ for the 8 freeform interfaces.
By comparing \autoref{fig:overview-accuracy} and \autoref{fig:perturb}, we can deduce that when the participants interacted with {\scshape{PAD\_16}}, the capacitance shift has an approximate strength of $\sigma = 2pF$.

\subsubsection{Mathematical Analysis}
To theoretically validate the hypothesis of the \singlewire connection being more robust, we perform a mathematical analysis.
To correctly recognize a touchpoint, $p$, we must satisfy two conditions:
\begin{align}
    \label{eq:next_touch}
    \lvert \TRecog{p} - \TCalib{p} \rvert & <  \lvert \TRecog{p} - \TCalib{p+1} \rvert \\
    \label{eq:prev_touch}
    \lvert \TRecog{p} - \TCalib{p} \rvert & <  \lvert \TRecog{p} - \TCalib{p-1} \rvert
\end{align}
These conditions avoid misrecognizing a touchpoint, $p$, with its two other adjacent touchpoints, $p-1$ and $p+1$.
Note that when $p=1$, we only need to satisfy \autoref{eq:next_touch}; similarly, when $p=N$, only \autoref{eq:prev_touch} is required. 

\subsubsection{Double-wire connection}
We let $c$ be the capacitance during the calibration stage and $(c+\epsilon)$ be the capacitance during the recognition stage. 
The goal is to derive the range of $\epsilon$ that satisfies both \autoref{eq:next_touch} and \autoref{eq:prev_touch}. 
By referring to \autoref{eq:time-two-wires-approx}, we can derive the range of $\epsilon$ as: 
\begin{equation}
    \label{eq:double_epsilon_range}
    - \frac{c}{2} \frac{r_{p}}{\RTill{p}} < \epsilon < \frac{c}{2} \frac{r_{p+1}}{\RTill{p}}
\end{equation}
The upper and lower bounds correspond to satisfying \autoref{eq:next_touch} and \autoref{eq:prev_touch}, respectively.

\paragraph{Single-wire connection}
We derive the range of $\epsilon$ using \autoref{eq:time-one-wire-approx}.
Unlike the \doublewire{}, the \singlewire{}'s logarithmic term depends on a touchpoint, $p$. 
To simplify the math expression, we denote the logarithmic term in \autoref{eq:time-one-wire-approx} as $\LogTerm{p}$.
To satisfy \autoref{eq:next_touch}, $\epsilon$ must satisfy:
\begin{equation}
    \label{eq:single_epsilon_next}
    \begin{cases}
    \epsilon < \frac{c}{2} \left( \frac{\RTill{p+1} \LogTerm{p+1}}{\RTill{p} \LogTerm{p}} - 1 \right), ~\mathrm{if}~ \RTill{p+1} \LogTerm{p+1} - \RTill{p}\LogTerm{p} > 0\\[5pt]
    \epsilon > -\frac{c}{2} \left( 1- \frac{\RTill{p+1} \LogTerm{p+1}}{\RTill{p} \LogTerm{p}} \right), ~\mathrm{otherwise} 
    \end{cases}
\end{equation}
Similarly, \autoref{eq:prev_touch} corresponds to:
\begin{equation}
    \label{eq:single_epsilon_prev}
    \begin{cases}
    \epsilon < \frac{c}{2} \left( \frac{\RTill{p-1} \LogTerm{p-1}}{\RTill{p} \LogTerm{p}} - 1 \right), ~\mathrm{if}~ \RTill{p-1} \LogTerm{p-1} - \RTill{p}\LogTerm{p} > 0\\[5pt]
    \epsilon > - \frac{c}{2} \left( 1 - \frac{\RTill{p-1} \LogTerm{p-1}}{\RTill{p} \LogTerm{p}} \right), ~\mathrm{otherwise} 
    \end{cases}
\end{equation}

When we apply our resistance optimization (\autoref{sec:one-wire-optimization}) to find $r_1$, $v_{N}(0)$ is only slightly smaller than $\VThres{}$ (e.g., \autoref{fig:single-wire}-b4).
In this case, we can assume $\TCalib{p+1} < \TCalib{p} < \TCalib{p-1}$. 
By concurrently considering this assumption with \autoref{eq:single_epsilon_next} and \autoref{eq:single_epsilon_prev}, we can derive:
\begin{equation}
    \label{eq:single_epsilon_range}
    -\frac{c}{2} \left( 1- \frac{\RTill{p+1} \LogTerm{p+1}}{\RTill{p} \LogTerm{p}} \right) < \epsilon < \frac{c}{2} \left( \frac{\RTill{p-1} \LogTerm{p-1}}{\RTill{p} \LogTerm{p}} - 1 \right)
\end{equation}

\paragraph{Comparison}
For the \doublewire{} connection, as indicated by \autoref{eq:double_epsilon_range}, we can make the range of $\epsilon$ larger (i.e., more robust to the capacitance change) by increasing $r_p$ and $r_{p+1}$ while keeping $\RTill{p}$ as small as possible. 
Also, since $\RTill{p}$ (i.e., $r_1 + \cdots + r_p$) increases as $p$ increases, generally, a larger $p$ is more difficult to achieve with a wider range of $\epsilon$.
Based on these observations, to create a more robust interface, the resistance values should be designed to have $r_1 < \cdots < r_N$ while ensuring a large difference between the adjacent resistance values.
For example, we can set $r_p = a r_{p-1}$ where $a > 1$. 
In this case, \autoref{eq:double_epsilon_range} becomes $- \frac{c (a^p - a^{p-1})}{2 (a^p - 1)} < \epsilon < \frac{c (a^{p+1} - a^{p})}{2 (a^p - 1)}$.
Ultimately, when $a \to \infty$, $-c/2 < \epsilon < \infty$.
However, \textit{in practice}, the minimum resistance ($r_1$) and the maximum resistance ($r_\nTouchpoints$) should be large and small enough, respectively.
These considerations are due to the limitations of the microcontroller's measurement of time delays, conduit volumes, and 3D printing resolutions.
For example, in a practical setting, we can set $\nTouchpoints = 10$ and $a=1.1$.
This configuration setting leads to $r_\nTouchpoints \approx 2.4 r_1$ and the average range of $\epsilon$  is $0.28c$ for touchpoints $p=\{2, \cdots, 9\}$ (i.e., $\frac{1}{8}\sum_{p=2}^{9} c \frac{r_{p+1} + r_{p}}{2\RTill{p}}$). 
Note that when we do not apply this optimization using $a > 1$, this range becomes smaller: e.g., $0.23c$ when $a=1$. 

For the \singlewire{} connection, by referring to \autoref{eq:single_epsilon_range}, we can infer that increasing the range of $\epsilon$ can be achieved by having relationships $r_2 > \cdots > r_N$ while keeping $r_1$ as small as possible. 
However, as discussed in \autoref{sec:one-vs-two}, $r_1$ must also satisfy $r_1 > (1 - \VThres{} / \VIn{}) \RTill{N}$.
To perform a fair comparison with the \doublewire{}, we set $\nTouchpoints = 10$, $r_{p-1} = 1.1 r_{p}$ for $p \geq 2$ (i.e., corresponding to $a=1.1$), and $r_1 = 1.01 (r_2 + \cdots r_\nTouchpoints)$.
Here we assume the use of Arduinno UNO R4 as a microcontroller, i.e., $\VIn{}{=}$\si{5\volt}, and $\VThres{}{=}$\si{{2.5}\volt}.
Note that $r_1$ is resistance of a resistor connected to a microcontroller and can be easily adjusted and large unlike the other resistance values.
Then, this setting derives $0.33c$ as the average $\epsilon$ range for touchpoints $p=\{2, \cdots, 9\}$.
When $r_{p-1} = r_{p}$ (i.e., a non-optimal case), this range becomes $0.28c$. % and $0.24c$ when $r_{p-1} = 0.9 r_{p}$.
% When $r_{p-1} = 1.2 r_{p}$, this range improves to $0.37c$.
These results support that the \singlewire{} connection can create more robust freeform interfaces than the \doublewire{} for our expected usage.

\section{Limitations and Future Work}
\label{sec:limitations}
This work introduces a computational design pipeline that embeds multiple capacitive touchpoints into any 3D model that has a closed mesh without self-intersection. Our method exploits RC Delay so that all touchpoints within our freeform interface can be capacitively sensed using only a \singlewire\ or \doublewire\ connection. Our six evaluations enable a thorough understanding of the RC Delay capacitive sensing technique, highlighting 
areas of improvement.

\subsection{Supporting Smaller Objects} 
Our fabrication scalability evaluation demonstrates that our approach for the \doublewire\ connection could potentially support embedding a touchpoint for every \si{9\milli\meter} distance.
The \singlewire\ connection places stricter constraints on fabricating a freeform interface with a smaller footprint (e.g., requiring over \si{30\milli\meter} distance between each pair of touchpoints). 
While our approach can generally support fabricating small objects (e.g., the smallest volume we fabricated is $\si{78284\milli\meter}^3$ for four touchpoints), future research is necessary on fabricating smaller objects (e.g., robotic grippers).

Our current fabrication scalability is largely dictated by the resistivity of the Protopasta conductive filament. 
Conductive filaments (including the Protopasta's) are typically used to connect electronic components (i.e., the role of wires), and are manufactured to have low resistivity.
In contrast, our approach uses the conductive filament to create 3D printed resistors, requiring a different need of electrical properties.
In our case, as long as the filament is conductive, a larger resistivity is generally preferable. A larger resistivity can help achieve the target resistance with a shorter conductive trace length.
Designing such a filament would make it possible to fabricate smaller freeform interfaces.

\subsection{Supporting More Distinct Signals} Our SNR evaluation highlights the robustness of our technique. All of the reported values in \autoref{tab:snr} are above the minimum SNR threshold ($> 7$) and achieve the standard for real-world applications ($> 15$)~\cite{davison2010techniques}. 
The \singlewire\ condition for Stanford Bunny and Hilbert Curve significantly outperforms the \doublewire\ condition.  
Though we see a drop in performance for the \scshape{PAD\_16} \normalfont for the \singlewire\ condition, this result also matches our discussion in \autoref{sec:one-wire} and insights from \autoref{sec4:scalability}. As expected, the \singlewire\ connection becomes more sensitive to noise if a freeform interface has too many touchpoints. This limitation is enforced by the resistance optimization discussed in \autoref{sec:one-wire-optimization}. One possible improvement could be relaxing the resistance value constraint we made for the efficient optimization (i.e., $r_2 = \cdots = r_N$). 
Optimizing each individual resistance value would create more distinct signals. 
However, we expect this approach would be subject to a much higher computational cost.
Similar to fabricating smaller objects, addressing this challenge requires producing higher resistance within a small volume. This could be achieved through the use of conductive filaments that have higher resistivity and shorter conductive trace lengths.

\subsection{Supporting Real-Time Calibration Adjustment} Though the SNR results highlight the robustness of our technique to background noise, our user study highlights a limitation of our pipeline. Currently, touchpoint selection is fully dependent on the calibration data. During the time gap between the calibration stage and touchpoint selection, if a change is introduced (e.g., a change in participant's capacitance or microcontroller performance), this dependency without real-time adjustment can introduce significant recognition errors. 

The \doublewire\ connection is more susceptible to errors due to this dependency.
In \autoref{sec:one-vs-two}, we originally hypothesized that the \doublewire\ would perform better given how we can generate numerous unique RC delays. In contrast, generating unique RC delays with the \singlewire\ connection is more restricted. However, our user study with \scshape{PAD\_16} \normalfont highlighted the trade-offs of the \doublewire\ connection in real-world conditions. As discussed in \autoref{sec8:robustness}, our computational experiment and mathematical analysis validate that if a change occurred after calibration, then the \doublewire\ connection has a more critical effect.
This calibration shift did not occur in the SNR tests given its short time duration (\si{{9}\sec}).
Thus, enabling more robust sensing will require future research on improving calibration, such as including a real-time adaptive baseline for adjustment. 

We imagine such solutions can include internally routing an additional wire that can gather baseline capacitance readings. Another solution is to use swept-frequency capacitive sensing~\cite{sato2012touche} with our technique. This combination can sweep different frequencies over a time window to sense whether the user configuration has changed. Regardless, future solutions would require the system to automatically detect the user’s capacitance and re-calibrate by referring to \autoref{eq:time-two-wires-approx} or \autoref{eq:time-one-wire-approx}.

\subsection{Supporting Multi-touch} Our approach demonstrated overall 91.42\% recognition accuracy (SE=1.329) for various objects across 10 participants. However, this technique is currently limited to one touch selection at a time. Future work should examine how to extend this technique to multi-touch. 
This will require modifying the algorithms in the automatic circuit design to account for multiple simultaneous touches. Given that RC delay values can be optimized with our approach, it is possible to design the traces such that the sum of each combination of RC delay values can also be a unique value. This approach is similar to creating a resistor ladder~\cite{chris_resistor_ladder} in which different combinations of resistors are used to uniquely identify multiple switches in a single circuit. Another interesting research direction would be to leverage machine learning to predict simultaneously activated touchpoints based on changes in the RC delay signal. 

\section{Conclusion}
\label{sec:conclusion}
We introduce a computational design pipeline that embeds multiple capacitive touchpoints into any 3D model that has a closed mesh without self-intersection. The core of our approach is optimizing a phenomenon called RC Delay so that all touchpoints within our freeform interface can be capacitively sensed using only a \singlewire\ or \doublewire\ connection. By leveraging multi-material printing, we achieve our research goal of streamlining fabricating interactive 3D printed objects with complex geometry with minimal instrumentation. The strengths of our pipeline (scalability, computational performance, robustness, accuracy, and applicability) work towards 3D printing objects that are fully interactive and ready for use straight off the printer as final products in real-world contexts.
\begin{acks}
This research is sponsored in part by the U.S. National Science Foundation through grants IIS-2040489, IIS-2320920, STEM+C 1933915, the Knut and Alice Wallenberg Foundation through Grant KAW 2019.0024, and the CU Boulder Engineering Education and AI-Augmented Learning Interdisciplinary Research Theme Seed Grant.
This work was also supported by an agreement with the National Renewable Energy Laboratory (NREL) under Alliance Partner University Program (APUP) No. UGA-0-41026-191. 
Most of the work was done while S. Sandra Bae and Takanori Fujiwara were respectively at CU Boulder and Link{\"o}ping University.
\end{acks}

\bibliographystyle{ACM-Reference-Format}
\bibliography{bib}
\appendix
\section{Equation Derivation with Symbolic Programming}
\label{appendix-a:equation-derivation}
We derived \autoref{eq:voltage-transient-two-wires} and \autoref{eq:voltage-transient-one-wire} by utilizing Lcapy~\cite{hayes2022lcapy}, a Python library that can perform symbolic circuit analysis. Symbolic circuit analysis can derive equations from given circuits and mathematical symbols (in our case, $v_\mathrm{in}$, $c$, $t$, $r_1, \cdots, r_{n+1}$)

\begin{figure*}[t!]
	\centering
	\captionsetup{farskip=0pt}
    \subfloat[Resistance measurement for horizontal traces]{
        \includegraphics[width=0.47\linewidth]{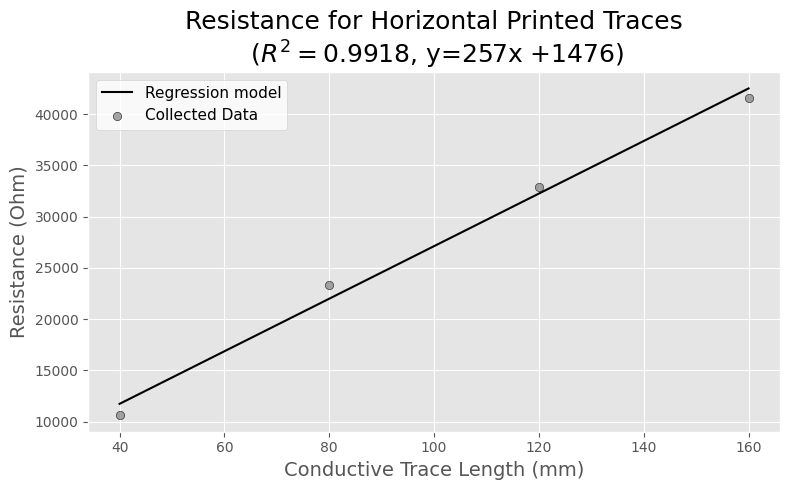}
        \label{fig:horizontal-resistance}
    }
    \hspace{3pt}
    \subfloat[Resistance measurement for vertical traces]{
        \includegraphics[width=0.47\linewidth]{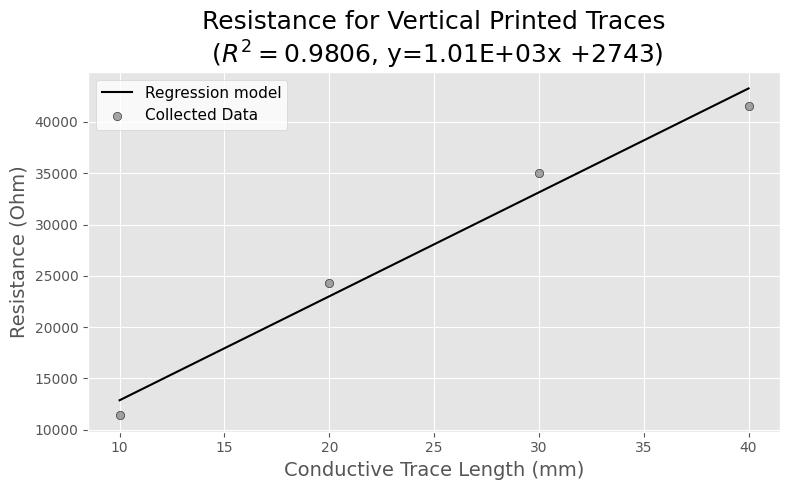}
        \label{fig:vertical-resistnace}
    }
    \caption{Two linear regression models for (a) horizontal traces and (b) vertical traces.}
	\label{fig:conductive-measurement}
\end{figure*}

\section{Implementation}
\label{appendix-b:implementation}

\subsection{Software Implementation}
The user interface to select the touchpoints in 
\autoref{fig:web-ui} is a web application made using three.js~\cite{threejs} and the three-mesh-bvh~\cite{three-mesh-bvh} libraries. 
We implemented the algorithms used in the automatic circuit design stage (\autoref{sec3:auto-circuit-design}) with Python 3 and libraries for matrix computations and machine learning methods such as NumPy/SciPy~\cite{virtanen2020scipy} and scikit-learn~\cite{pedregosa2011scikit}. 
We used graph-tool~\cite{graphtool2014peixoto} to use algorithms such as Dijkstra's and A* for path finding.
We used PyVista~\cite{sullivan2019pyvista} (a Python API for Visualization Toolkit~\cite{vtkBook}) for 3D graphics-related operations such as clipping, voxelization, and ray tracing.
For the resistance optimization for the single-wire condition, we used Lcapy~\cite{hayes2022lcapy} and SymPy~\cite{meurer2017sympy} for the circuit simulation and symbolic computation and Pathos for multiprocessing.
For the calibration, we utilized the sensing-network library~\cite{bae2023computational}.

\begin{table}[h!]
\renewcommand{\arraystretch}{0.98}
  \small
  \centering
  \begin{tabular}{>{\raggedright}p{3cm}  
                 ccccc}

\toprule
    \multirow{2}{*}{Conductive Trace} \\[3pt] Length (\si{\milli\meter}) & ~ &  \multirow{2}{*}{S1 (\si{\ohm})} & \multirow{2}{*}{S2 (\si{\ohm})} & \multirow{2}{*}{S3 (\si{\ohm})} & \multirow{2}{*}{Avg (\si{\ohm})}\\
     \arrayrulecolor{black!30}\midrule

    %% ------------------------------------------------------
    %% Horizontal
    %% ------------------------------------------------------

    \textbf{Horizontal} &
    ~ &
    ~ &
    ~ & 
    ~ \\\addlinespace[-0.2em]
    
    \quad 40 & ~ &
    10600 &
    10630 & 
    10600 & 
    10610
    \\

    \quad 80 & ~ &
    23330 &
    23630 & 
    23110 & 
    23357
    \\

    \quad 120 & ~ &
    33290 &
    33020 & 
    32470 & 
    32927
    \\

    \quad 160 & ~ &
    40200 &
    43070 & 
    41600 & 
    41623
    \\

    \midrule

  %% ------------------------------------------------------
    %% Vertical
    %% ------------------------------------------------------

    \textbf{Vertical} &
    ~ &
    ~ &
    ~ & 
    ~ \\\addlinespace[-0.2em]

    \quad 10 & ~ &
    10730 &
    12500 & 
    10970 & 
    11400
    \\

    \quad 20 & ~ &
    26070 &
    22670 & 
    24080 & 
    24274
    \\

    \quad 30 & ~ &
    37700 &
    38830 & 
    28540 & 
    35023
    \\

    \quad 40 & ~ &
    43830 &
    42600 & 
    38330 & 
    41587
    \\
\arrayrulecolor{black}\bottomrule

\end{tabular}
  \caption{Resistance values of conductive traces. Three samples (S1-S3) for each measurement. For the thickness of the conductive trace, we followed our computational design pipeline default (i.e., horizontal: \si{{0.8}\milli\meter}, vertical: \si{{1.2}\milli\meter}).}
  \label{tab:measurements}
\end{table}

\subsection{3D Printing Hardware and Materials}
To fabricate freeform interfaces with embedded multi-points within one pass, we rely on a multi-material FDM 3D printer using non-conductive and conductive filaments.
We use a Snapmaker J1S 3D printer, which supports dual-nozzle printing. Both nozzles are \SI{0.4}{\milli\meter} standard soft brass.  We set our print speed to \SI{60}{\milli\meter/\second} for both conductive and non-conductive filaments. The layer height of all prints is \SI{0.24}{\milli\meter}.

Our conductive filament is Protopasta’s conductive PLA (1.75 mm)~\cite{protopasta}. This filament is commonly available and provides a good balance of conductivity and resistivity to design a sensing network. The non-conductive filament can be any standard PLA filament. The print and build plate temperatures for both filaments used were based on vendor recommendations.

The infill percentage differs for the four files discussed in \autoref{sec3:multi-material}. The original body uses an infill of 20\% using the gyroid pattern. In some of our preliminary tests, we found that not having enough infill (e.g., 0\%) can cause parasitic capacitance~\cite{en12122278} where the coupled charge dissipates due to the air inside the model's body. As such, we recommend choosing a range between 5--20\% infill to provide steady sensor readouts. The other STL files all have an infill percentage of 100\% using the rectilinear pattern.

\subsection{Sensing System}
For our microcontroller, we tested using the Arduino Uno R4 WiFi (48MHz CPU), which has a 5V power source and a 2.5V logic threshold. To constantly measure the time delays, we utilized digital signals from the microcontroller’s digital I/O pins.

\section{Measurement of Conductive Trace's Resistance}
\label{appendix-c:resistance-test}

We conducted an empirical investigation to understand the resistivity properties of the conductive filament. We measured the conductive traces horizontally and vertically as our 3D printed conductive traces are drawn in a serpentine trace pattern. We produced three samples of each measurement, and we used Fluke 115 Digital Multimeteter to measure each object's resistance (\autoref{tab:measurements}). See the supplemental materials to see the STL files.

In \autoref{fig:conductive-measurement}, we used the average values in (\autoref{tab:measurements}) to plot the relationship between the conductive trace's length and measured resistance. The two linear regression models informed how to generate the conductive traces during the circuit embedding step (\autoref{sec:circuit-embedding}).

\end{document}